\documentclass[11pt,a4paper]{article}
\usepackage[T1]{fontenc}
\usepackage[utf8]{inputenc}
\usepackage{amsmath,amsthm,newtxtext,newtxmath}
\usepackage[margin=25mm,headheight=14pt]{geometry}
\usepackage{graphicx,booktabs,longtable,array,pdflscape}
\usepackage[round,authoryear]{natbib}
\usepackage{algorithm,algpseudocode}
\usepackage{enumitem,microtype,placeins}
\usepackage[font=small,labelfont=bf]{caption}
\usepackage[colorlinks=true,linkcolor=blue,citecolor=blue,urlcolor=blue]{hyperref}
\newtheorem{assumption}{Assumption}
\newtheorem{theorem}{Theorem}
\newtheorem{proposition}{Proposition}
\newtheorem{lemma}{Lemma}
\newtheorem{corollary}{Corollary}
\theoremstyle{definition}
\newtheorem{definition}{Definition}
\theoremstyle{remark}
\newtheorem{remark}{Remark}
\newcommand{\E}{\mathbb E}
\newcommand{\Prob}{\mathbb P}
\newcommand{\R}{\mathbb R}
\newcommand{\ind}{\mathbf 1}
\newcommand{\trans}{^{\mathsf T}}
\newcommand{\norm}[1]{\lVert #1\rVert}
\newcommand{\doi}[1]{\href{https://doi.org/#1}{\nolinkurl{https://doi.org/#1}}}
\DeclareMathOperator{\col}{col}

\DeclareMathOperator{\RSS}{RSS}

\setlist[enumerate]{itemsep=3pt,topsep=5pt}
\setlist[itemize]{itemsep=3pt,topsep=5pt}
\hypersetup{pdftitle={Selective inference for pruned interaction trees with known unequal variances in randomized trials},pdfauthor={}}
\title{Selective inference for pruned interaction trees\\ with known unequal variances in randomized trials}
\author{Tomoshige Nakamura \thanks{Faculty of Health Data Science, Juntendo University}}
\date{1 November 2026}
\begin{document}
\maketitle
\vspace{-1.5em}
\begin{abstract}
Interaction trees reuse outcomes to discover treatment-effect heterogeneity, complicating inference after growth and pruning. We develop selective inference for generalized least-squares (GLS) interaction trees with a common covariate basis and known unequal variances. Under a conditional Gaussian model with positive diagonal covariance, weighted orthogonality reduces survival under additive fixed-penalty pruning to one quadratic inequality supplementing the growth event. This yields exact finite-sample conditional tests and confidence intervals, with retained-path and complete-tree inference agreeing on compatible sections. Under causal identification and a correct reference prognosis, the pruning-invariant GLS target is a signed finite-design weighted contrast of conditional average treatment effects. Simulations at $n=100,200,400,800$ and maximum depths one to three search all adjacent-observation midpoints, support calibration and demonstrate over-rejection when pruning survival is omitted. Compared with a 50:50 honest procedure, full-data learning improves reference-split recovery in some small-sample settings. Root-level median intervals can be shorter, while deep-node intervals and upper tails can be much wider; these comparisons concern each procedure's selected targets. Estimating a common scale with the variance ratio fixed produces modest changes in Selective coverage in the settings examined, without extending the exact guarantee. An exploratory ACTG 175 analysis provides a working-model illustration with estimated arm variances.

\end{abstract}
\section{Introduction}\label{sec:introduction}
Randomized controlled trials (RCTs) establish how treatments compare on average, but clinical decisions also require understanding whether their relative benefits vary across patients. A treatment that is useful overall may be more beneficial for patients with particular combinations of baseline characteristics. Discovering such subgroups can generate hypotheses for subsequent trials and help describe treatment-effect variation \citep{Su2011BCEI,Athey2016}. The relevant question concerns differences between treatment responses: a characteristic that predicts a good outcome under both treatments need not identify patients who benefit more from one treatment than the other.

Interaction trees offer an interpretable approach to this problem. They recursively partition participants using treatment-by-subgroup comparisons, producing baseline rules that describe possible effect modifiers \citep{Su2009}. Their clinical applications include exploration of quality-of-life responses to a psychoeducational programme for breast cancer survivors \citep{Su2011BCEI} and substance-use treatment engagement following motivational interviewing \citep{Steingrimsson2019}. An ensemble extension, random forests of interaction trees, has been used to study variation in the response to acupuncture for chronic headache \citep{Su2018RFIT}. These applications motivate the subgroup question addressed here; the present theory is developed for a specified single-outcome tree, with longitudinal and ensemble extensions requiring separate arguments.

After a tree has been discovered, an additional question arises: how strong is the statistical evidence for its apparent treatment-effect differences? The same outcomes usually determine the split variables, thresholds, and which branches remain after pruning. An ordinary interaction test at a selected split therefore treats an outcome-dependent choice as if it had been specified in advance. This reuse of outcomes is closely related to the model-selection problem studied in selective inference for the lasso, where inference on selected coefficients accounts for the data-dependent choice of model \citep{Lee2016}. Selective inference for lasso-selected effect modification has also been developed using residualization and asymptotic arguments \citep{Zhao2022}. Existing tree-based work provides inferential tools, including an overall interaction permutation test \citep{Su2011BCEI}, bootstrap-calibrated subgroup intervals \citep{Loh2019}, and uncertainty estimates for interaction-tree ensembles \citep{Su2018RFIT}. The issue addressed here is more specific: finite-sample conditional tests and confidence intervals for individual retained treatment interactions after growing and pruning a specified deterministic tree on the same data.

Tree-Values provides a route to such inference for regression trees \citep{Neufeld2022}. Building on conditional Gaussian selective inference \citep{Lee2016}, it accounts for the CART selection event when testing response means or contrasts in discovered regions. Under its Gaussian model with known common variance, it controls the conditional type-I error of an individual selected test and provides confidence intervals with conditional coverage. Its theory includes fixed-penalty pruning. The paper also discusses general known-covariance Gaussian conditioning, while noting that the computational reductions need not extend. The Box Lunch Study application, involving a regression analysis of caloric intake with estimated variance, gives a concrete illustration of inference for CART-discovered groups; it is not a treatment-heterogeneity analysis. Extending this approach to treatment interactions requires new structural arguments because their contrasts compare treated and control outcomes within each child region, rather than only comparing child outcome means.

Sample splitting supplies another established route to inference after subgroup discovery. Honest causal trees learn subgroups using one sample and estimate their effects using independent observations \citep{Athey2016}. Independence protects the estimation stage from the outcome-driven search. The allocation of observations to the two stages can nevertheless affect the learned tree, the resulting inferential target, and interval precision. Although the overall analysis uses both parts, each stage has access to only a subset of the responses. This constraint can be consequential when the trial is small, or when recursive splits leave few treated and control participants in a parent region. Using all responses for construction and inference avoids the additional random partition, but requires accounting for selection and need not produce shorter intervals. The trade-off between independent estimation and selective reuse is therefore an empirical question as well as a theoretical one.

We extend the selective-inference strategy of Tree-Values to a covariate-adjusted interaction tree for quantitative RCT outcomes. The tree uses generalized least-squares (GLS) interaction scores, a prespecified prognostic basis common to every node, and additive pruning with a fixed penalty. The Gaussian covariance is known and diagonal, allowing different outcome variances in the treatment arms. Our main contribution is to establish the weighted orthogonality that makes this extension tractable: along the appropriate target-specific perturbation, on a section preserving the grown path, off-path candidates and other selected scores remain fixed. This yields an exact pruning-survival constraint, finite-sample conditional tests and confidence intervals, and a factorization showing that retained-path conditioning and complete-tree conditioning give identical inference on compatible sections. The pruning objective and Gaussian conditioning principle have precedents; the new results concern their validity and computational reduction for these residualized treatment interactions. We also identify the inferential target and its causal interpretation. It is a finite-design GLS interaction coefficient and, under causal identification and a correctly specified reference-treatment prognosis, a signed weighted contrast of conditional average treatment effects. It generally differs from an ordinary subgroup-average-effect difference; within-child homogeneous effects recover the child-to-child difference. Equal outcome variance is an explicit special case.

The experiments assess both the inferential correction and its practical cost at sample sizes $n=100,200,400,800$ and maximum growth depths one to three. We search all adjacent-observation midpoints and compare naive inference, growth-only selective inference, pruning-aware inference, and a same-rule 50:50 honest procedure. The comparisons distinguish split-level error from tree-level error, and calibration from the ability to form a tree, infer a retained interaction, and achieve useful precision. Stronger pruning exposes the additional correction needed after the growth event. Full-data learning can improve recovery of a reference split in a small sample, while root-level median intervals can be shorter and their upper tails longer than those of the honest comparator. The targets and selected populations generally differ, and neither discovery rates nor interval widths establish a uniform advantage. An additional experiment estimates the common scale while keeping the variance ratio known. An ACTG 175 application in HIV infection compares two combination regimens using original-scale 20-week CD4 counts and an unpruned tree with maximum depth three. Its estimated arm variances, residual tail departures, and public-subset limitations make it a working-model illustration; the exact guarantee requires the stated Gaussian model and known covariance. These restrictions also distinguish the projection-level inference developed here from an unrestricted claim about clinical subgroup benefits.

The remainder of the article is organized as follows. Section~\ref{pr:sec:setup} introduces the RCT setting, interaction trees, and the outcome-reuse problem. Section~\ref{sec:proposed} presents the proposed selective-inference procedure, including the interaction-tree construction, target contrasts, and selection-adjusted tests and intervals. Section~\ref{sec:theoretical} establishes the target properties and selective-inference results. Section~\ref{sec:simulation} reports the simulation design, comparison procedures, and results for small and moderate samples, Section~\ref{sec:actg} presents the ACTG 175 analysis, and Section~\ref{sec:discussion} discusses limitations and further work. Appendices~\ref{supp:sec:proofs} and~\ref{supp:sec:extensions} contain the proofs and additional structural results.

\section{Problem setup and interaction trees}
\label{pr:sec:setup}

\subsection{RCT setting and the subgroup question}
For participants $i=1,\ldots,n$, let $X_i\in\R^p$ be baseline covariates, $A_i\in\{0,1\}$ the assigned treatment, and $Y_i$ a quantitative outcome. Write $Y_i(a)$ for the potential outcome under treatment $a$. The scientific question is whether combinations of baseline characteristics describe variation in the relative treatment response, rather than merely predicting a high or low outcome under either treatment. The scientific target is variation in the conditional average treatment effect (CATE), $\tau(x)=\E\{Y_i(1)-Y_i(0)\mid X_i=x\}$. Such variation can motivate interpretable candidate subgroups for further investigation.

For the statistical analysis, condition throughout on the realized design $\mathcal D=(X,A)$. Fix a known positive diagonal matrix $V$, specified conditional on $\mathcal D$ without using $Y$. The exact selective-inference results use the following sampling model; randomization alone does not imply it.

\begin{assumption}[Gaussian sampling with known diagonal covariance]
\label{pr:ass:gaussian}
Conditional on $\mathcal D$, $Y\sim N_n(\mu,\sigma^2V)$ for arbitrary $\mu\in\R^n$, known $0<\sigma^2<\infty$, and the known positive diagonal $V$ above. For arm-specific variances this includes
\begin{equation}\label{gls:eq:arm-variance}
\sigma^2=\sigma_0^2,\qquad
V_{ii}=\ind\{A_i=0\}+\rho\ind\{A_i=1\},\qquad
\rho=\sigma_1^2/\sigma_0^2>0,
\end{equation}
where both arm variances are known. Conditional independence of outcomes is retained; correlated outcomes are not covered by the local structural results below.
\end{assumption}
Known variance is also part of the exact Gaussian theory in Tree-Values \citep{Neufeld2022}. Randomization does not imply equality of the two potential-outcome variances. Here it is the known variance shape, rather than equality across arms, that is used in weighted construction and conditioning.

For causal interpretation, let $B_{i\cdot}=b(X_i)\trans$ for a fixed finite-dimensional baseline basis $b(x)\in\R^q$ containing an intercept. The same basis, restricted to the participants in a region, will be used at every node, and its specification uses no outcomes. The next assumption separates causal identification in the analyzed sample from the additional restriction on the reference-treatment prognosis. The latter is not supplied by randomization. This assumption is needed for the causal reading of the target defined in Section~\ref{sec:gls-scores}, but not for Gaussian inference on that projection target.

\begin{assumption}[Causal identification and reference prognosis]
\label{pr:ass:causal}
Potential outcomes satisfy consistency and no interference. Conditional on the full covariate design $X$, the allocation vector $A$ is independent of the full potential-outcome vector, and $0<\Prob(A_i=1\mid X)<1$. For $a=0,1$, suppose $\E\{Y_i(a)\mid X\}=m_a(X_i)$ and $m_0(x)=b(x)\trans\beta$ for an unknown $\beta$. Write $\tau(x)=m_1(x)-m_0(x)$, so that
\begin{equation}\label{pr:eq:causal-mean}
\mu=B\beta+A\odot\tau_X,\qquad \tau_X=(\tau(X_1),\ldots,\tau(X_n))\trans.
\end{equation}
These assumptions concern the analyzed sample. Selection into an available data file is not supplied with an identification guarantee by randomization in an original, larger trial.
\end{assumption}
Under Assumption~\ref{pr:ass:causal}, $m_1(x)-m_0(x)$ agrees with this CATE by iterated expectation. The vector $\mu$ is the mean before conditioning on the subsequently selected tree.

\subsection{Interaction trees}
An interaction tree recursively divides the baseline covariate space to explore treatment-effect variation \citep{Su2009}. The root contains all analyzed participants. At a reached parent region $R$, a candidate rule $s$, such as a baseline-variable threshold, divides $R$ into a left child $R_L$ and a right child $R_R$. Repeating this operation produces a sequence of baseline rules for each terminal group.

The distinction from an ordinary outcome-prediction tree is the comparison used to assess a split. To illustrate it, suppose all four treatment-by-child cells are nonempty and write $\bar Y_{a,C}$ for the observed mean in treatment arm $a$ and child $C\in\{L,R\}$. An unadjusted empirical interaction contrast for a fixed candidate is
\[
\widehat\Delta_s
=(\bar Y_{1,L}-\bar Y_{0,L})
 -(\bar Y_{1,R}-\bar Y_{0,R}).
\]
It compares the observed treatment differences across the two children. A split can have little difference between the children's overall outcome means while exhibiting a large treatment interaction, or conversely. Interaction-tree splitting uses a measure of this interaction, commonly standardized for uncertainty, to compare candidate rules. Covariate-adjusted interaction trees additionally account for prognostic baseline variables \citep{Steingrimsson2019}.

This four-cell display explains the interaction principle. It is not the definition of the adjusted GLS score or projection target used in this paper, and conditional on a realized design it is not asserted to be an unbiased estimate of a pure causal subgroup-effect difference. Section~\ref{sec:gls-scores} specifies our common-basis GLS construction.

Growing a tree applies the split comparison recursively within the resulting children. A stopping rule limits further growth, for example through a maximum depth or a minimum number of observations in each treatment-by-child cell. Pruning then collapses branches of a grown tree to obtain a smaller collection of interactions. Interaction-complexity pruning balances the retained split scores against a penalty per retained split \citep{Su2009,Steingrimsson2019}. The specific deterministic growth rule, fixed-penalty objective, and tie convention analyzed here are specified in Section~\ref{sec:gls-scores}, under Assumptions~\ref{pr:ass:growth} and~\ref{pr:ass:pruning}; the results do not cover every implementation called an interaction tree.

\subsection{Outcome reuse and the inferential problem}
The observed outcomes enter both stages of this search: they determine which candidate rule wins at each reached node and which grown splits survive pruning. An interaction selected for having a large score is therefore not a prespecified contrast. The ordinary reference distribution for testing a fixed interaction does not automatically remain appropriate after selecting it from competing rules. Even when the numerical pruning penalty is fixed in advance, the event that a particular branch survives still depends on the outcomes.

The question in this paper is how to test and form a confidence interval for the interaction at a prespecified node address when it is retained, using the same responses for tree construction and inference. A collapsed or ungrown address has no split-specific inferential result to report. The target is the adjusted treatment-by-child projection defined in Section~\ref{sec:gls-scores}; its causal interpretation and distinction from an ordinary subgroup-average-effect difference are established in Section~\ref{sec:target-properties}. The reporting goal is individual selected-split inference. A guarantee for every retained split separately is not, by itself, a familywise guarantee for the whole tree.

Tree-Values addresses outcome reuse for CART response-mean targets by conditioning on selection \citep{Neufeld2022}. We use the same selective-inference strategy for the specified interaction tree. Section~\ref{sec:proposed} defines the procedure, and Section~\ref{sec:theoretical} derives the restrictions imposed by growth and pruning and the resulting conditional law. Those restrictions, rather than an ordinary fixed-contrast reference distribution, determine the test and confidence interval.

\section{Proposed selective inference procedure}\label{sec:proposed}
The proposed method uses all observed responses to construct the specified interaction tree and to perform inference on one retained treatment interaction. Its inputs are the observed design and outcomes, an outcome-independent common basis and growth rules, maximum depth, known $\sigma$ and $V$, fixed penalty $\lambda$, a prespecified node address, null value $\theta_0$, and level $\alpha$. If that address is grown and retained, the output is its original GLS point estimate, a selection-adjusted two-sided $p$-value, and a $1-\alpha$ confidence interval. Otherwise the output is unavailable.

The procedure has three stages: identify the selected split and its original interaction contrast; compute all statistic values that reproduce its ancestral growth choices and survival after pruning; and use the Gaussian distribution restricted to that set for testing and interval inversion. It retains the same point estimate as the full-data naive procedure, while changing the reference distribution to account for selection. Section~\ref{sec:gls-scores} specifies the tree and contrasts; Section~\ref{sec:algorithm} gives the proposed inference. Section~\ref{sec:theoretical} establishes its conditional guarantees.

\subsection{Interaction-tree construction and target contrasts}
\label{sec:gls-scores}
\label{pr:sec:growth}
\label{pr:sec:pruning}
Use the common basis $B$ and known covariance shape $V$ introduced in Section~\ref{pr:sec:setup}, and write $W=V^{-1}$. Rank deficiency in the nuisance basis is allowed; $P_C$ denotes the unique Euclidean orthogonal projection onto $\col(C)$. Each node uses the restrictions $B_R$, $V_R$, and $W_R$ of these same objects. The construction below specifies the actual adjusted interaction used for growth and inference.

In a region $R$, a candidate split $s$ has left indicator $h_s$ and right indicator $1-h_s$. Restrict all vectors in the next display to $I_R=\{i:X_i\in R\}$ and define
\begin{equation}\label{pr:eq:contrast}
\begin{split}
C_s&=[B_R,A_R,h_s],\qquad g_s=A_R\odot h_s,\\
r_s&=(I-P_{V_R^{-1/2}C_s})V_R^{-1/2}g_s,\\
v_s&=V_R^{1/2}r_s,\qquad d_s^2=r_s\trans r_s=v_s\trans W_Rv_s,\\
\eta_s&=W_Rv_s/d_s^2,\qquad \ell_s=W_Rv_s/d_s,\qquad
u_s=V_R^{-1/2}v_s/d_s.
\end{split}
\end{equation}
The residual $v_s$ is unique even if the nuisance coefficients are not: it minimizes $(g_s-C_sa)\trans W_R(g_s-C_sa)$ over $a$. Only candidates with $d_s>0$ are retained. Extend all candidate vectors by zero outside $I_R$. The vector $u_s$ is unit length in whitened coordinates; $\ell_s$ is the score loading in the original outcome coordinates and satisfies $\ell_s\trans V\ell_s=1$. Their score, estimate, and finite-design target are
\begin{equation}\label{pr:eq:score-target}
G_s(y)=(\ell_s\trans y)^2
=\frac{(v_s\trans Wy)^2}{d_s^2},\qquad
T_s=\eta_s\trans Y,\qquad \theta_s=\eta_s\trans\mu.
\end{equation}
The global products use the zero extensions and global $W$. Both outcome and model columns must be weighted; scaling the outcomes alone is not this GLS rule.

For a fixed eligible candidate, $\theta_s$ is the coefficient of the interaction column $g_s$ in the weighted least-squares approximation of the mean vector:
\[
\min_{a,t}
(\mu_R-C_sa-g_st)\trans W_R(\mu_R-C_sa-g_st).
\]
All quantities in this objective are restricted to the parent $I_R$. Here $C_s$ contains the prognostic basis and the treatment and child main effects, while $g_s=A_R\odot h_s$ is the local treatment-by-child interaction. Write $v_{s,R}$ and $\eta_{s,R}$ for the restrictions of the zero-extended vectors to this parent. The residual identities $v_{s,R}\trans W_RC_s=0$ and $v_{s,R}\trans W_Rg_s=d_s^2$ give the weighted Frisch--Waugh coefficient
\[
\theta_s=\frac{v_{s,R}\trans W_R\mu_R}{d_s^2}
=\eta_{s,R}\trans\mu_R=\eta_s\trans\mu.
\]
Replacing $\mu_R$ by the observed $Y_R$ gives the corresponding GLS point estimate $T_s$ in~\eqref{pr:eq:score-target}. The eligibility condition $d_s>0$ means that $g_s\notin\col(C_s)$, so this interaction coefficient is unique even if the nuisance coefficients $a$ are not. This is a finite-design projection target; it does not assume that the mean vector lies exactly in $\col([C_s,g_s])$.

The parameter $\theta_s$ is fixed for a fixed candidate and design, but the reported candidate is selected from the outcomes. Its causal interpretation requires the additional conditions in Assumption~\ref{pr:ass:causal} and is established in Section~\ref{sec:target-properties}.

\paragraph{Deterministic growth.}

\begin{assumption}[Finite deterministic local growth]
\label{pr:ass:growth}
Fix a finite maximum depth $D$, with the root at depth zero. At each reached region of depth less than $D$, a finite ordered candidate set, eligibility, and stopping rules depend only on $\mathcal D$ and the region. Each eligible node is split by maximizing $G_s(y)$. Candidates with identical matrices $\ell_s\ell_s\trans$ are represented by the first in the fixed priority order, retaining its geometric split. Other ties use that same order. A node at depth $D$ or with no eligible candidates is terminal. There is no outcome-dependent stopping, basis selection, or pruning \emph{during this growth stage}.

The result $\widehat H(y)$ is a complete labeled tree: its labels include split variables, thresholds, and child directions, and not merely topology. Every node uses the common basis in Section~\ref{pr:sec:setup}.
\end{assumption}

A design-feasible labeled grown tree obeys these eligibility and stopping rules with eligible prescribed split labels. This term does not assert that its score-winning event has positive probability. It also covers the forced completions defined below.

\paragraph{Fixed-penalty pruning.}

A rooted pruned subtree $P\preceq H$ is obtained only by collapsing some internal nodes and all their descendants into terminal nodes. Equivalently, $I(P)$ is an ancestor-closed subset of $I(H)$. The root-only tree, with $I(P)=\varnothing$, is allowed. Retaining a split retains both immediate child regions, which can themselves be terminal.

\begin{assumption}[Additive interaction pruning with collapse at equality]
\label{pr:ass:pruning}
Fix $0\le\lambda<\infty$ without using $Y$. More generally, $\lambda$ may be fixed conditional on $\mathcal D$, $V$, and the known $\sigma$. On the grown tree $H=\widehat H(y)$, maximize
\begin{equation}\label{pr:eq:objective}
Q_\lambda(P;y)=\sum_{w\in I(P)}G_w(y)-\lambda|I(P)|,
\qquad P\preceq H.
\end{equation}
Select the maximizer with the fewest internal nodes by the following explicit recursion. At terminal nodes set $F_w^H(y)=0$. At internal nodes set
\begin{equation}\label{pr:eq:dp}
V_w^H(y)=G_w(y)-\lambda+F_{w_L}^H(y)+F_{w_R}^H(y),
\qquad F_w^H(y)=\max\{0,V_w^H(y)\}.
\end{equation}
Starting at the root, retain $w$ if $V_w^H(y)>0$ and then apply the same rule to both children; otherwise collapse $w$. In particular, equality is resolved by collapsing. Denote the result by $\widehat P_\lambda(y)$.

No branch is raised or moved. The local contrasts of retained splits are not redefined. The parameter, criterion, and reporting rule are not tuned by the same outcomes.
\end{assumption}

The penalty in~\eqref{pr:eq:objective} has weighted squared-outcome units for the specified normalization of $V$. Using dimensionless scores $G_w/\sigma^2$ and a fixed dimensionless penalty $\widetilde\lambda$ is equivalent to setting $\lambda=\sigma^2\widetilde\lambda$. Substitution of an estimated $\sigma$ is not included in this equivalence under the exact known-variance guarantee.

\begin{proposition}[Optimization and exact tie convention]
\label{pr:prop:optimizer}
The recursion in Assumption~\ref{pr:ass:pruning} returns the unique maximizer of~\eqref{pr:eq:objective} having the fewest internal nodes. Its optimal value is $F_{\rm root}^H(y)$. Once the grown tree and scores are available, this pruning takes a number of arithmetic operations linear in the number of nodes.
\end{proposition}
\begin{remark}[Equal-variance special case]
\label{gls:rem:equal-variance}
When $V=I_n$, the definitions in~\eqref{pr:eq:contrast} give $r_s=v_s$, $u_s=\ell_s=v_s/d_s$, and $\eta_s=v_s/d_s^2$. With the same basis, candidates, stopping rules, priorities, penalty, and known scale, the scores, grown and pruned trees, original targets, and selection sections coincide pointwise with those of the same interaction-tree procedure using unweighted residuals. Consequently its selective $p$-values and confidence intervals coincide wherever defined. The comparison is with an unweighted interaction tree, rather than CART response-mean contrasts. In~\eqref{gls:eq:arm-variance}, this is $\rho=1$; the common variance $\sigma^2$ need not equal one.
\end{remark}

\subsection{Proposed selective tests and confidence intervals}\label{sec:algorithm}
At a prespecified address that is observed to be retained, record its selected ancestral path $\gamma$ and use its original contrast $\eta$, with estimate $t_{\rm obs}=\eta\trans Y$ and target $\theta=\eta\trans\mu$. For this calculation, the recorded path labels and hence its contrast are held fixed, while their selection event is imposed. Use the covariance-matched decomposition in~\eqref{pr:eq:decomposition}: set $c=V\eta/(\eta\trans V\eta)=v$, hold $z=Y-ct_{\rm obs}$ fixed, and vary the responses along $y_z(t)=z+ct$. The known Gaussian reference standard deviation of $T=\eta\trans Y$ is $\omega=\sigma/d$.

The growth section $\mathcal T_\gamma^{\rm grow}(z)$ contains every $t$ for which the recorded path wins all its target and ancestor comparisons. The proposed procedure uses the further restriction
\[
J(z)=\mathcal T_\gamma^{\rm grow}(z)
\cap\{t:d^2t^2>\delta_\gamma(z)\},
\]
where the second set is the exact pruning-survival gate in~\eqref{pr:eq:ret-section}. Its threshold is computed from the path's ancestor scores and hanging-subtree pruning values in~\eqref{pr:eq:a-target}--\eqref{pr:eq:delta}. The computation includes every component of $J(z)$. Growth-only inference uses the growth section alone; at positive penalties it can omit restrictions imposed by reporting only pruning survivors. Naive inference instead uses the unrestricted Gaussian reference distribution.

Let $F^J_{a,\omega}$ denote the CDF of $N(a,\omega^2)$ restricted to $J$, as defined in~\eqref{pr:eq:cdf}. We evaluate $U=F^{J(z)}_{\theta_0,\omega}(t_{\rm obs})$, report $p=2\min\{U,1-U\}$, and invert this same CDF for the equal-tail confidence interval using~\eqref{pr:eq:endpoints}. Thus the point estimate remains $t_{\rm obs}$, while both the test and interval reflect growth and pruning selection. Theorem~\ref{pr:thm:pivot} gives their individual conditional validity under the stated known-covariance Gaussian model; the causal interpretation requires the additional conditions in Section~\ref{sec:target-properties}. Conditional validity does not guarantee a short interval or simultaneous error control for the whole tree.

Algorithm~\ref{alg:selective} summarizes the complete procedure for one prespecified address. A collapsed or ungrown address has no split-specific inferential result to report. The procedure retains the original local contrast and does not refit an effect parameter using the final terminal leaves.

\begin{algorithm}[tbp]
\caption{Selective inference for one retained treatment interaction}\label{alg:selective}
\begin{algorithmic}[1]
\Require $\mathcal D,Y$, fixed common basis, growth rules and maximum depth, known $\sigma,V$, fixed penalty $\lambda$, prespecified address, null value $\theta_0$, level $\alpha$
\State Compute the common weighted candidate contrasts~\eqref{pr:eq:contrast}; grow $\widehat H(Y)$ with the corresponding GLS scores.
\State Compute~\eqref{pr:eq:dp} bottom-up; retain nodes top-down only when their continuation values are strictly positive.
\If{the specified address is not internal in $\widehat P_\lambda(Y)$}
  \State \Return unavailable.
\EndIf
\State Retrieve its original target path $\gamma$, $\eta$, $v$, $\ell$, and $d$.
\State Set $t_{\rm obs}=\eta\trans Y$, $c=v$, $z=Y-vt_{\rm obs}$, and $\omega=\sigma/d$.
\State Construct the entire growth section $J=\mathcal T_\gamma^{\rm grow}(z)$ from all target and ancestor comparisons in~\eqref{pr:eq:quadratic}.
\State Compute hanging-subtree pruning values and $\delta_\gamma(z)$ using~\eqref{pr:eq:a-target}--\eqref{pr:eq:delta} on $H_\gamma(z)=\widehat H(Y)$.
\State Set $J\gets J\cap\{t:d^2t^2>\delta_\gamma(z)\}$, retaining every component.
\State Evaluate $U=F^J_{\theta_0,\omega}(t_{\rm obs})$ and $p=2\min\{U,1-U\}$.
\State Solve~\eqref{pr:eq:endpoints} for the confidence endpoints.
\State \Return $\widehat\theta=t_{\rm obs}$, $p$, and the confidence interval.
\end{algorithmic}
\end{algorithm}

Only hanging-subtree values and selected ancestor scores enter $\delta_\gamma(z)$. Their values at $Y$ equal their values at $z$, so cached non-target scores can be reused; unconstrained regrowth at $z$ would generally select a different path and is unnecessary. The growth section for a given path can also be reused across fixed penalties and deeper growth prefixes that preserve the path, while the pruning threshold must be recomputed for each depth and penalty.

The implementation uses $S=\ell\trans Y=dT$, with variance $\sigma^2$. Its perturbation direction is $b=v/d$, so $Y-bS=Z$ and $\ell\trans b=1$. In general $b\ne\ell$: perturbing along the score loading would give the wrong Gaussian section under unequal variances. On the $S$ scale, the pruning gate is $S^2>\delta_\gamma(Z)$, and interval endpoints for its mean are divided by $d$ to obtain the interval for $\theta$. Equivalently, whitening outcomes and all candidate columns gives the unit direction $u=V^{-1/2}v/d$. At $V=I_n$, $b=\ell=u$ and the original implementation is recovered.

All positive-length support components are used in probability calculations. Gaussian masses are evaluated on a log scale using lower-tail or survival probabilities, and mean inversion uses Corollary~\ref{pr:cor:endpoints}. Floating-point tolerances for near-zero residual norms, slopes, and inversion are distinct from the mathematical exactness statements. The fixed $V$ and scale $\sigma$ are inputs, not variance estimates learned from the analyzed outcomes. Numerical candidate reductions and implementation tolerances are documented in Section~\ref{supp:sec:implementation} of Appendix~\ref{supp:sec:extensions}.

\section{Theoretical analysis}\label{sec:theoretical}
This section establishes the validity and computational structure of the proposed procedure. We first establish the causal interpretation of the GLS target and its invariance under pruning. We then establish weighted orthogonality, which enables us to derive the exact selection event imposed by tree growth and pruning. These results yield the truncated-Gaussian law and the conditional guarantees for the proposed tests and confidence intervals under the known-covariance Gaussian model. Finally, we show that retained-path and full-tree conditioning give identical inference on compatible sections. Complete proofs are provided in Appendix~\ref{supp:sec:proofs}.

\subsection{Causal interpretation and pruning invariance}
\label{sec:target-properties}
\begin{proposition}[GLS target, causal interpretation, and pruning invariance]
\label{pr:prop:causal}
Under Assumption~\ref{pr:ass:causal}, for each fixed eligible split,
\begin{equation}\label{pr:eq:weighted-target}
\theta_s=\sum_{i\in I_R:A_i=1}\eta_{s,i}\tau(X_i).
\end{equation}
The sums of $\eta_s$ in the left-treated, left-control, right-treated, and right-control cells are, respectively, $1,-1,-1,1$. If $\tau$ is constant at $\tau_L$ and $\tau_R$ at the realized design points of the two children, then $\theta_s=\tau_L-\tau_R$. In particular, a constant effect on the parent implies $\theta_s=0$.

Pruning by collapsing descendants leaves this target unchanged for a retained split, provided its original local contrast in~\eqref{pr:eq:contrast} is used. Its parent and immediate child partition must not be redefined.
\end{proposition}
The weights depend on the known variance shape as well as allocation and covariate imbalance, and can be signed. For the same geometric split, the GLS and OLS targets generally differ; both equal $\tau_L-\tau_R$ under the within-child homogeneity in Proposition~\ref{pr:prop:causal}. Thus the target is generally not a difference of ordinary subgroup-average effects. Zero contrast does not imply effect homogeneity. If the reference prognosis is $B\beta+r$ instead, the projection target additionally contains $\eta_s\trans r$. Gaussian projection coverage then has no automatic interpretation as coverage for a pure treatment-effect contrast. Re-estimation of leaf effects after pruning is a separate inferential problem.

\subsection{Weighted orthogonality}
\begin{lemma}[Weighted descendant orthogonality and orthonormal scores]
\label{pr:lem:orthogonality}
For a split $s$ in region $R$, let $D'$ be one of its children or any region contained in that child. Every candidate $r$ in $D'$ satisfies
\begin{equation}\label{gls:eq:orthogonality}
v_s\trans Wv_r=0,\qquad
\ell_s\trans V\ell_r=u_s\trans u_r=0.
\end{equation}
Contrasts on disjoint parent regions satisfy the same identities. Consequently, for any feasible fixed labeled grown tree $H$, its whitened selected vectors $\{u_w:w\in I(H)\}$ are orthonormal and orthogonal to $V^{-1/2}C_0$, where $C_0=[B,A]$. For these fixed tree labels, the linear scores $S_w=\ell_w\trans Y$ have variance $\sigma^2$ and zero pairwise covariance under Assumption~\ref{pr:ass:gaussian} before conditioning on tree selection.
\end{lemma}
Unlike a CART sibling-mean contrast, $\ell_s$ need not be constant within a child. The common-basis restriction argument proves~\eqref{gls:eq:orthogonality} for these adjusted interaction directions. It is this weighted local geometry that permits the section simplification below, rather than a general claim for arbitrary non-diagonal covariance.
\subsection{Retained-path event and truncation set}
\label{pr:sec:selection}

Fix a feasible labeled path $\gamma$ \emph{as an index before conditioning on its selection}. It consists of internal node addresses $w_0,\ldots,w_k$, their prescribed split labels $s_0,\ldots,s_k$, and the directions followed to the next node. Its target is $\star=w_k$. Feasibility means every prescribed split is eligible in its design-determined region and $k<D$. Let $\eta=\eta_\gamma$, $\ell=\ell_\gamma$, $v=v_\gamma$, and $d=d_\gamma$ be its target contrast quantities. Define
\begin{equation}\label{pr:eq:decomposition}
\begin{aligned}
T&=\eta\trans Y,\qquad
c=\frac{V\eta}{\eta\trans V\eta}=v,\\
Z&=Y-vT,\qquad
\omega^2=\sigma^2\eta\trans V\eta=\sigma^2/d^2.
\end{aligned}
\end{equation}
The quantities $\eta,\ell,v,c,d,\omega$ are fixed for the fixed design and path index; $T$ and $Z$ remain random. Set $y_z(t)=z+ct$ for $z\in\eta^\perp$.

In a region $R$, let $r\prec s$ mean that $r$ has higher priority than $s$. Its exact winning event is
\begin{equation}\label{pr:eq:win}
\mathcal W_{R,s}=
\bigcap_{r\prec s}\{G_s(Y)>G_r(Y)\}
\ \cap\!
\bigcap_{s\prec r}\{G_s(Y)\ge G_r(Y)\}.
\end{equation}
The growth-path event and its section are
\begin{equation}\label{pr:eq:grow-event}
E_\gamma^{\rm grow}=\bigcap_{j=0}^{k}\mathcal W_{R_j,s_j},
\qquad \mathcal T_\gamma^{\rm grow}(z)
=\{t:y_z(t)\in E_\gamma^{\rm grow}\}.
\end{equation}
The intersection is exact: the root winner fixes its child region, then the next winner fixes the next, and so on. Off-path decisions do not affect growth along this path.

\begin{definition}[Canonical grown and pruned completions]
\label{pr:def:completion}
For every $z\in\eta^\perp$, construct $H_\gamma(z)$ by forcing the prescribed splits along $\gamma$, and growing every other reached node using $z$, with the original candidates, priorities, and stopping rules.

Within this fixed canonical tree, prune every hanging off-path subtree using~\eqref{pr:eq:dp} evaluated at $z$. Force all path splits, including the target, to remain. Denote the resulting tree by $P_\gamma(z)$. The hanging subtrees are the target's two children and the sibling branches that leave its ancestral path.

These constructions are defined even when $\mathcal T_\gamma^{\rm grow}(z)$ is empty. They do \emph{not} mean growing an unconstrained tree at outcome $z$, which could choose different path splits.
\end{definition}

\begin{lemma}[Canonical completion along a growth section]
\label{pr:lem:completion}
Under Assumption~\ref{pr:ass:growth}, for every $z\in\eta^\perp$ and $t\in\mathcal T_\gamma^{\rm grow}(z)$,
\begin{equation}\label{pr:eq:completion}
\begin{aligned}
\widehat H(y_z(t))&=H_\gamma(z),\\
G_\star(y_z(t))&=d^2t^2,\\
G_w(y_z(t))&=G_w(z)\quad (w\in I(H_\gamma(z)),\ w\ne\star).
\end{aligned}
\end{equation}
The completion maps are measurable.
\end{lemma}
For the next definitions, all pruning values are computed \emph{on $H_\gamma(z)$}, not on $\widehat H(z)$. Write
\[
F_w^\gamma(z)=F_w^{H_\gamma(z)}(z).
\]
For $j<k$, let $b_j$ be the sibling of $w_{j+1}$, which hangs off the path at $w_j$. Set
\begin{align}
a_k(z)&=-\lambda+F_{w_{k,L}}^\gamma(z)+F_{w_{k,R}}^\gamma(z),\label{pr:eq:a-target}\\
a_j(z)&=a_{j+1}(z)+G_{w_j}(z)-\lambda+F_{b_j}^\gamma(z),
\qquad j=k-1,\ldots,0,\label{pr:eq:a-ancestor}\\
\delta_\gamma(z)&=\max_{0\le j\le k}\{-a_j(z)\}.\label{pr:eq:delta}
\end{align}
Only hanging-subtree values and prescribed ancestor scores occur in these expressions. They are functions of $z$ alone.

\begin{theorem}[Exact pruning-survival threshold and section]
\label{pr:thm:gate}
Under Assumptions~\ref{pr:ass:growth} and~\ref{pr:ass:pruning}, let $E_\gamma^{\rm ret}$ be the event that the prescribed path, including its target split as an internal node, occurs in $\widehat P_\lambda(Y)$. Then, with the exact tie rules,
\begin{equation}\label{pr:eq:ret-event}
E_\gamma^{\rm ret}=E_\gamma^{\rm grow}
\cap\{d^2T^2>\delta_\gamma(Z)\}.
\end{equation}
For every $z\in\eta^\perp$, its exact section is
\begin{equation}\label{pr:eq:ret-section}
\boxed{\quad
\mathcal T_{\gamma,\lambda}^{\rm ret}(z)
=\mathcal T_\gamma^{\rm grow}(z)
\cap\{t:d^2t^2>\delta_\gamma(z)\}.
\quad}
\end{equation}
The pruning gate is all of $\R$ when $\delta_\gamma(z)<0$, $\R\setminus\{0\}$ when it equals zero, and the two open tails outside $[-\sqrt{\delta_\gamma(z)}/d,\sqrt{\delta_\gamma(z)}/d]$ when it is positive.
\end{theorem}
\begin{proposition}[Finite computation of the complete section]
\label{pr:prop:geometry}
For each selected-candidate/competitor comparison on the fixed path, set $a_r=\ell_r\trans z$ and $b_r=\ell_r\trans c$. Its exact strict or weak comparison is the corresponding version of
\begin{equation}\label{pr:eq:quadratic}
\{a_s-a_r+(b_s-b_r)t\}\{a_s+a_r+(b_s+b_r)t\}\ge0.
\end{equation}
If $m_\gamma$ is the total number of these comparisons, the retained section is a finite union of intervals, possibly with isolated points, and has at most $2m_\gamma+2$ finite boundary points from nonconstant comparison polynomials and the pruning gate. All components must be retained for inference. Corollary~\ref{gls:cor:two-components} in Appendix~\ref{supp:sec:extensions} sharpens this general bound to at most two interval pieces for the matching-GLS rule.

At an observed $y$ whose target survives, $H_\gamma(z_{\rm obs})=\widehat H(y)$. Thus the grown tree and its cached scores can be reused; no unconstrained regrowth at $z_{\rm obs}$ is needed. Once the scores are available, a bottom-up pass gives the hanging-subtree values, and a pass along the path computes $\delta_\gamma(z_{\rm obs})$.
\end{proposition}
\subsection{Selective tests and confidence intervals}
\label{pr:sec:inference}

For a measurable set $J\subseteq\R$ with positive Lebesgue measure, define the truncated-normal CDF
\begin{equation}\label{pr:eq:cdf}
F^J_{a,\omega}(t)
=\frac{\displaystyle\int_{J\cap(-\infty,t]}\phi_\omega(w-a)\,dw}
{\displaystyle\int_J\phi_\omega(w-a)\,dw},
\qquad
\phi_\omega(x)=\frac{1}{\sqrt{2\pi}\,\omega}
\exp\left(-\frac{x^2}{2\omega^2}\right).
\end{equation}
All integrals use the entire section, including every interval component. Endpoints or isolated points do not change these integrals.

\begin{theorem}[Known-variance inference conditional on a retained path]
\label{pr:thm:pivot}
Under Assumptions~\ref{pr:ass:gaussian}, \ref{pr:ass:growth}, and~\ref{pr:ass:pruning}, fix a path index $\gamma$ with $\Prob(E_\gamma^{\rm ret}\mid\mathcal D)>0$ and set $\theta=\eta\trans\mu$. For almost every $z$ under its selected distribution,
\begin{equation}\label{pr:eq:law}
T\mid(Z=z,E_\gamma^{\rm ret},\mathcal D)
\sim N(\theta,\omega^2)\text{ restricted to }
\mathcal T_{\gamma,\lambda}^{\rm ret}(z).
\end{equation}
In particular, the section has positive normal probability almost surely on the event, and
\begin{equation}\label{pr:eq:pivot}
F^{\mathcal T_{\gamma,\lambda}^{\rm ret}(Z)}_{\theta,\omega}(T)
\mid(E_\gamma^{\rm ret},\mathcal D)\sim\operatorname{Unif}(0,1).
\end{equation}
For a fixed null value $\theta_0$, the two-sided equal-tail $p$-value is
\begin{equation}\label{pr:eq:pvalue}
p_{\theta_0}
=2\min\left\{
F^{\mathcal T_{\gamma,\lambda}^{\rm ret}(Z)}_{\theta_0,\omega}(T),
1-F^{\mathcal T_{\gamma,\lambda}^{\rm ret}(Z)}_{\theta_0,\omega}(T)
\right\}.
\end{equation}
Under $\theta=\theta_0$, its conditional rejection probability at $0<\alpha<1$ is exactly $\alpha$. The confidence set
\begin{equation}\label{pr:eq:ci}
\mathcal C_{\gamma,\lambda}(Y)
=\left\{a\in\R:
\frac\alpha2\le
F^{\mathcal T_{\gamma,\lambda}^{\rm ret}(Z)}_{a,\omega}(T)
\le1-\frac\alpha2\right\}
\end{equation}
has conditional coverage $1-\alpha$ for $\theta$. Both guarantees also hold conditional on $Z$ for almost every selected $z$.
\end{theorem}
\begin{corollary}[Unique finite equal-tail interval endpoints]
\label{pr:cor:endpoints}
Under Theorem~\ref{pr:thm:pivot}, almost surely on $E_\gamma^{\rm ret}$ the confidence set is a closed interval $[L,U]$ with unique finite endpoints solving
\begin{equation}\label{pr:eq:endpoints}
F^{\mathcal T_{\gamma,\lambda}^{\rm ret}(Z)}_{L,\omega}(T)=1-\alpha/2,
\qquad
F^{\mathcal T_{\gamma,\lambda}^{\rm ret}(Z)}_{U,\omega}(T)=\alpha/2.
\end{equation}
No bound on realized width or finite expected width is asserted.
\end{corollary}
The guarantee is finite-sample under the stated Gaussian model with known covariance. Small $n$ affects eligibility, the residual information $d^2$ and the realized interval width, but does not add an asymptotic sample-size requirement to the conditional pivot.
\begin{remark}[Ties and the unpruned limit]
\label{pr:rem:ties}
All event and section identities above use exact growth priorities and collapse at pruning equality. They hold pointwise. If strict inequalities are replaced by weak inequalities for numerical integration, equality boundaries do not change the normal probabilities for almost every selected $z$. At growth nodes, nonidentical $\ell_s\ell_s\trans$ give nonzero comparison polynomials in the full Gaussian outcome; finitely many possible labels imply a null union of their tie sets. The pruning boundary has at most two points for each $z$ because $d>0$. Fubini's theorem handles exceptional fibers on which a growth comparison is identically zero.

At $\lambda=0$, every selected gain is strictly positive almost surely under Assumption~\ref{pr:ass:gaussian}, by the same finite-label argument. The entire grown tree is then retained, and the unpruned inference is recovered almost surely. Exact ties at individual outcomes remain governed by the stated collapse convention.
\end{remark}

\begin{corollary}[A prespecified node address and availability]
\label{pr:cor:address}
Fix a node address before observing $Y$, and let $\mathcal V$ be the event that it is an internal node of the final pruned tree. If $\Prob(\mathcal V\mid\mathcal D)>0$, the interval for its selected target has coverage $1-\alpha$ conditional on $(\mathcal V,\mathcal D)$.
\end{corollary}
This is an individual-node guarantee. Without further adjustment, selecting the reported address by its smallest $p$-value, largest estimate, or narrowest interval adds a selection event. Multiple-node simultaneous coverage is not supplied by this corollary. A collapsed node is unavailable for this split-specific inference; failure to retain a split does not establish homogeneous treatment effects.

\subsection{Equivalence with full-tree conditioning}
\label{pr:sec:factorization}

\begin{lemma}[Canonical final tree on a retained section]
\label{pr:lem:prunedcompletion}
Under the growth and pruning assumptions, for every $z$ and $t\in\mathcal T_{\gamma,\lambda}^{\rm ret}(z)$,
\begin{equation}\label{pr:eq:prunedcompletion}
\widehat H(y_z(t))=H_\gamma(z),\qquad
\widehat P_\lambda(y_z(t))=P_\gamma(z).
\end{equation}
\end{lemma}
\begin{theorem}[Exact factorization and compatible-section equivalence]
\label{pr:thm:fulltree}
Assume~\ref{pr:ass:growth} and~\ref{pr:ass:pruning}. Fix a complete labeled final tree $P$ containing $\gamma$ as an internal path. Let
\[
E_P=\{\widehat P_\lambda(Y)=P\},\qquad
\mathcal A_P^\gamma=\{z\in\eta^\perp:P_\gamma(z)=P\}.
\]
Then
\begin{equation}\label{pr:eq:factor-final}
\boxed{\quad E_P=E_\gamma^{\rm ret}\cap\{Z\in\mathcal A_P^\gamma\}.\quad}
\end{equation}
For a feasible complete grown tree $H$ with $P\preceq H$, define
\[
E_{H,P}=\{\widehat H(Y)=H,\ \widehat P_\lambda(Y)=P\},\qquad
\mathcal A_{H,P}^\gamma
=\{z:H_\gamma(z)=H,\ P_\gamma(z)=P\}.
\]
There is likewise the exact factorization
\begin{equation}\label{pr:eq:factor-pair}
E_{H,P}=E_\gamma^{\rm ret}\cap\{Z\in\mathcal A_{H,P}^\gamma\}.
\end{equation}
All sets in these factorizations are measurable. Thus, for $K=P$ or $K=(H,P)$, the full-event section satisfies, for every $z$,
\begin{equation}\label{pr:eq:section-case}
\mathcal T_K(z):=\{t:y_z(t)\in E_K\}
=\begin{cases}
\mathcal T_{\gamma,\lambda}^{\rm ret}(z),&z\in\mathcal A_K^\gamma,\\
\varnothing,&z\notin\mathcal A_K^\gamma.
\end{cases}
\end{equation}
In particular, at every compatible observed $y\in E_K$ and $z=y-c\eta\trans y$,
\begin{equation}\label{pr:eq:equal-sections}
\mathcal T_K(z)=\mathcal T_{\gamma,\lambda}^{\rm ret}(z).
\end{equation}

If Assumption~\ref{pr:ass:gaussian} also holds and the relevant event has positive probability, the retained-path pivot, test, and confidence interval are exact conditional on the complete final tree alone, or on the pair of complete grown and final trees. At the observed compatible section they coincide with inference computed directly from those full events.
\end{theorem}
\begin{remark}[What is, and is not, being equated]
\label{pr:rem:equivalence}
The comparison is between \emph{growth plus survival of the target path} and the complete final-tree event. Growth-only conditioning omits the pruning gate and generally gives different inference. Equality is asserted at compatible sections; a full-tree section is empty at an incompatible $z$ even when the retained-path section is nonempty.

Off-path quantities are not absent from the procedure: $\delta_\gamma(z)$ uses hanging-subtree pruning values. They can be calculated once after $Z$ is fixed and add no further $t$-dependent tree choices on the retained section. Ancestor \emph{competitors} need not be orthogonal to the target, so all ancestor growth comparisons remain in~\eqref{pr:eq:grow-event}. No independence between the target and the off-path tree \emph{after selection} is asserted.

The equality yields a computational reduction in section construction. It does not make intervals shorter than their complete-tree-conditioned counterparts, which are identical, and pruning does not guarantee useful interval width.
\end{remark}

\begin{remark}[A simultaneous reporting option]
\label{pr:rem:simultaneous}
At one prespecified penalty $\lambda$, fix a complete final tree $P$ of positive selection probability and put $K=|I(P)|$. Under the assumptions of Theorem~\ref{pr:thm:fulltree}, testing its projection nulls $H_{0,v}:\eta_v\trans\mu=0$ at level $\alpha/K$ controls the familywise error probability conditional on $(E_P,\mathcal D)$ at at most $\alpha$. Likewise, intervals of individual level $1-\alpha/K$ cover all $K$ selected targets jointly with conditional probability at least $1-\alpha$. If $K=0$, no tests or intervals are reported. These conclusions follow by the union bound after conditioning on $P$, where $K$ and the null indices are fixed; the target-specific orthogonal components need not coincide. This is a within-tree option, not a simultaneous guarantee across penalties or variance shapes. The simulations distinguish unadjusted $\alpha=0.05$ \emph{per retained node} from Bonferroni-adjusted tree-level rejection. Reported interval widths use nominal 95\% individual intervals; widths of simultaneous intervals are not evaluated. A justification is provided in Section~\ref{supp:proof:simultaneous} of Appendix~\ref{supp:sec:proofs}.
\end{remark}

The present theory allows unequal known arm variances and, more generally, a known positive diagonal shape, but requires known scale and a penalty fixed before observing the outcomes. Appendix~\ref{supp:sec:extensions} gives an additive fixed-family extension, including certain fixed leaf-count constraints (Proposition~\ref{pr:prop:fixed-family}), and counterexamples showing why neither arbitrary pruning nor omission of target survival is covered automatically. Cross-validated or manually outcome-selected penalties, different pruning risks, outcome-selected bases, and variance substitution require additional arguments. The causal interpretation remains conditional on Assumption~\ref{pr:ass:causal}; the Gaussian projection guarantee alone is not population subgroup-average-effect inference.

\section{Simulation}\label{sec:simulation}
We investigate calibration after growing and pruning an interaction tree, the additional correction required by pruning, and the practical consequences of reusing a small trial for discovery and inference. The experiments also compare a same-rule honest procedure, examine the upper tail of interval widths, and assess substitution of an estimated common variance scale.

\subsection{Data-generating mechanism}\label{sec:sim-design}
For each participant $i=1,\ldots,n$, we independently generate
\[
X_{ij}\overset{\mathrm{iid}}\sim\operatorname{Unif}(-1,1),\quad j=1,\ldots,5,
\qquad A_i\sim\operatorname{Bernoulli}(1/2),
\qquad \varepsilon_i\sim N(0,1),
\]
with $X_i$, $A_i$ and $\varepsilon_i$ mutually independent, and set
\begin{equation}\label{eq:sim-model}
Y_i=m_0(X_i)+A_i\tau(X_i)
       +\sqrt{1+(\rho-1)A_i}\,\varepsilon_i,
\qquad m_0(x)=2x_3+x_4-x_5.
\end{equation}
Thus the control and treatment error variances are $1$ and $\rho$, respectively. We use $n\in\{100,200,400,800\}$ and $\rho\in\{1/4,1,4\}$. The common reference basis is $b(x)=(1,x_1,\ldots,x_5)\trans$, of dimension $q=6$. The prognosis $m_0$ lies in its span and is annihilated by every eligible interaction contrast; this design evaluates selection adjustment under correct prognostic specification, rather than sensitivity to prognostic misspecification.

We adapt the three-level mean-tree construction of \citet[Section~5.1]{Neufeld2022} to treatment-effect heterogeneity:
\begin{equation}\label{eq:sim-effect}
\tau(x)=0.5+\gamma\ind\{x_1\le0\}
\left[1+a\ind\{x_2>0\}+\ind\{x_2x_3>0\}\right].
\end{equation}
There are five parameter settings,
\[
(\gamma,a)\in\{(0,1),(0.5,1),(1,1),(2,1),(1,0.5)\}.
\]
At $\gamma=0$, the CATE is the constant $0.5$ and every interaction projection target is zero. This is a null of no mean-effect modification, compatible with unequal arm variances. The value of $a$ is then immaterial, so this setting is generated only once. Positive $\gamma$ changes the amplitude of the three-level structure; $a=0.5$ weakens the intermediate split relative to its descendants. Changing $a$ also changes the root contrast. A reference representation splits first on $X_1$, then on $X_2$ within $X_1\le0$, and on $X_3$ in both resulting children. This representation need not be unique. All non-null settings have a depth-three structure, so maximum depths one and two restrict how much of it can be represented.

\subsection{Tree construction and comparison procedures}\label{sec:sim-methods}
At each reached parent, candidate generation enumerates every midpoint between distinct adjacent observed values of each covariate in that parent. The true cut point zero is not added to the search. Fixed feature and cut-point priorities resolve ties. Each growing split must have at least eight observations in every treatment-by-child cell and an identifiable weighted interaction. Numerical eligibility thresholds and representative-candidate reductions are specified in Section~\ref{supp:sec:implementation} of Appendix~\ref{supp:sec:extensions}. All nodes use the common basis and the known variance shape $V_{ii}=1+(\rho-1)A_i$. Growth has no outcome-dependent stopping. We consider maximum depths $D=1,2,3$, with root depth zero, and fixed additive subtree penalties $\lambda=0,4,8$, collapsing at equality. Depth $D$ permits at most $D$ splits along a path; it does not force that depth to be reached. We grow to depth three once, take each unpruned depth prefix, and apply pruning separately to that prefix. Since the true common scale is $\sigma=1$, the numerical penalties equal their variance-standardized values.

The four procedures differ as follows.
\begin{enumerate}[label=(\roman*)]
\item \emph{Naive} uses the full-data grown and pruned tree and an ordinary known-variance Gaussian reference distribution, without selection adjustment.
\item \emph{Growth-only} uses the same tree and targets, conditions on the complete ancestral growth path, and omits the pruning-survival constraint.
\item \emph{Selective} adds the pruning-survival constraint to the growth event, using the procedure of Section~\ref{sec:proposed}.
\item \emph{Honest} randomly divides the trial into equal training and holdout samples. The training sample grows and prunes a GLS tree with the same basis, candidate-generation policy, minimum-cell rule and numerical penalty. Its learned thresholds are applied unchanged to the independent holdout sample, where ordinary GLS inference is performed. This implements the sample-separation principle of \citet{Athey2016}, rather than reproducing a different causal-tree splitting objective.
\end{enumerate}
Honest inference requires at least one holdout observation in each treatment-by-child cell, at least $q+4=10$ holdout observations in the parent, and an identifiable interaction. The first three procedures share the full-data survivor set, GLS estimates and targets. Honest inference has its own learned tree, eligibility and holdout-design target. Its error-rate denominator therefore generally differs from theirs.

These rules impose concrete feasibility limits. A growing parent needs at least 32 observations, and a path with three internal splits needs at least $4\times16=64$. Consequently, the honest training sample of 50 at $n=100$ cannot produce a third-stage split, although shallower splits can be inferred. This is a property of the specified 50:50, minimum-eight comparator. The comparison does not assert an advantage over every possible honest design.

The principal experiment uses the true covariance. An additional module replaces only the common reference scale by an estimate, retaining the same known variance ratio and numerical penalty. Its construction and results are reported in Section~\ref{sec:sim-plugin}.

\subsection{Performance measures and Monte Carlo design}\label{sec:sim-evaluation}
For each sample size, we generate 5,000 independent constant-effect trials and 1,000 independent trials for each of the four non-null settings: 36,000 independent base trials overall. Within a base trial, the covariates, allocation, unit Gaussian errors and honest partition are shared across the three variance ratios, producing 108,000 correlated outcome versions. Each version is reused across the three depth limits and three penalties. Outcome versions, nodes and derived settings are not independent Monte Carlo replications. A separate 480-trial pilot checked computation and retention before the penalties, replication counts and endpoint subset were frozen; pilot results are not pooled with the main experiment.

Every potential node address is recorded, including ungrown, pruned and holdout-ineligible positions. Inference availability is the probability that at least one target can be computed, using all trials, including empty trees. Splitwise rejection at $p<0.05$ and coverage condition on successfully computed eligible retained splits. The denominator is the total number of such splits, not the number of participants or trials, and is reported separately for the full-data and honest procedures. Coverage is evaluated against the actual $\theta_v=\eta_v\trans\mu$ for each selected analysis design. The piecewise CATE values in \eqref{eq:sim-effect} are not substituted for these projections. Rejection is a false-positive rate (FPR) at $\gamma=0$; under heterogeneity it describes discovery among selected targets, rather than power for a common fixed parameter.

To assess recovery, we adapt the split-matching strategy of \citet[Section~5.4]{Neufeld2022}. Both procedures are evaluated on the same fixed set of 4,096 independently generated covariate vectors. Within each reference parent, its binary child labels are compared with an estimated split's three labels: left child, right child, or outside the estimated parent. A reference split is matched to the retained estimated split with largest adjusted Rand index (ARI); ARI greater than $0.75$ defines recovery. Deterministic traversal order resolves ties, and multiple reference splits may match one estimated split. We report recovery and the joint event of recovery, successful inference and rejection, each over all trials. This is local partition matching, not recovery of an entire hierarchical tree. In particular, a shallow estimated split can agree with a deeper reference split when labels are restricted to the reference parent. Actual growth depth is assessed separately.

All eligible trials evaluate the true-target pivot and its coverage decision. Confidence-interval endpoints and widths are computed only for the prespecified trial indices 1--500 in each sample-size/effect/ratio setting, without replacing unavailable intervals. We retain every computed width and summarize its count, median, 90th percentile and extremes. Width groups with fewer than 30 intervals are flagged. Fixed-address and trial-wise events use Wilson Monte Carlo intervals. Pooled depth and tree rates use ratio estimates with trial-cluster Monte Carlo standard errors (MCSEs), retaining trials with no available splits. Paired differences are used for the same-target full-data procedures and for known versus estimated scale within a procedure. General Honest comparisons concern different selected populations and targets. Reported Monte Carlo intervals are pointwise across the correlated reporting grid.

Individual tests are distinguished from the probability of any rejection in a tree. The latter uses all trials, including empty trees. We also apply Bonferroni at $0.05/K$, where $K$ is the method's retained split count, including holdout-ineligible honest splits, and make no rejection for an empty tree. These comparisons are within fixed configurations, with no selection of a penalty from the main outcomes.

Independent numerical checks compared candidate scores and coefficients with direct weighted QR and GLS calculations, checked selection sections against actual regrowth and pruning, and verified the known/estimated-scale target identities. All main records passed post-run completeness and pairing checks; there were no inference or endpoint failures. Extreme intervals were reconstructed and their endpoints checked independently at 100-decimal precision. The accompanying code specifies the study settings, seeds and commands needed to regenerate the results.

\subsection{Calibration and the need to adjust for pruning}\label{sec:sim-calibration}
\begin{figure}[htbp]
\centering
\includegraphics[width=\linewidth]{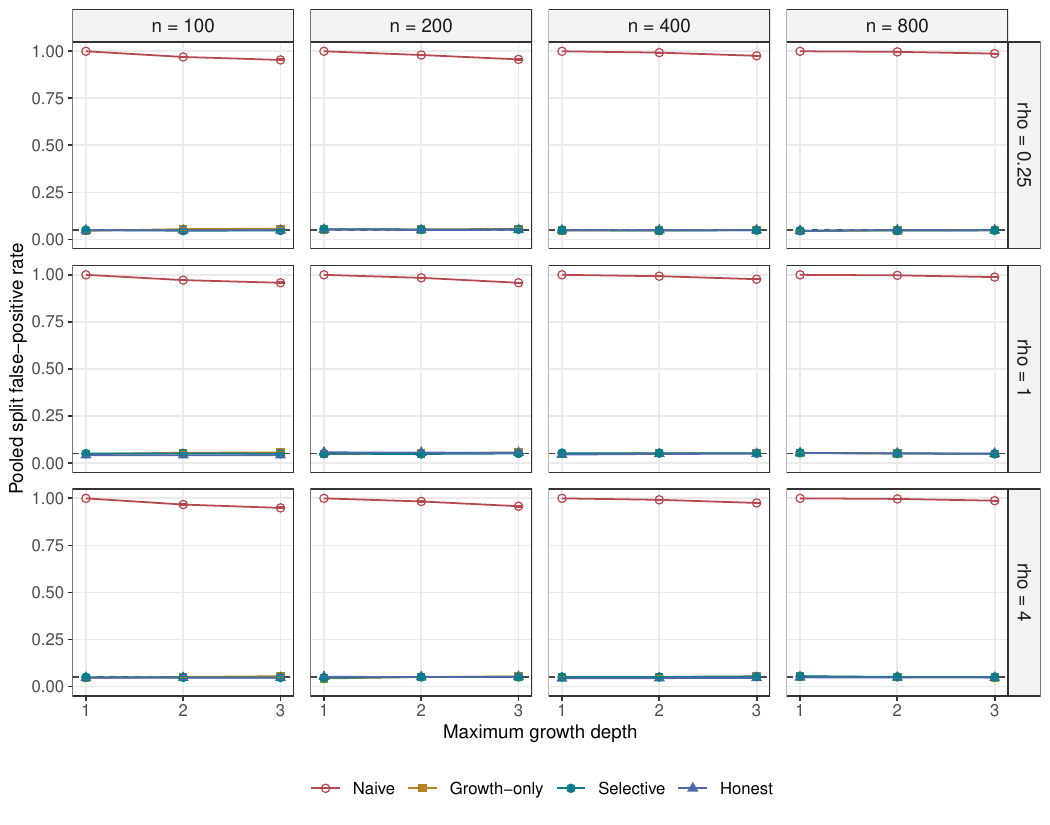}
\caption{Constant-effect FPR across maximum growth depths $D=1,2,3$, with fixed $\lambda=4$. Rows specify variance ratio and columns sample size. Rates pool successfully computed retained splits, with each method's own denominator; the three full-data procedures share their targets. Bars are pointwise trial-cluster Monte Carlo intervals and the dashed line is $0.05$. These are individual-split rates, not tree-wide error probabilities.}\label{fig:dense-calibration}
\end{figure}

Figure~\ref{fig:dense-calibration} compares the three growth limits at $\lambda=4$, and Table~\ref{tab:dense-calibration} gives both methods' split counts and the stronger-pruning comparison at $\rho=1$, $D=3$. In the constant-effect model at $D=3$, $\lambda=4$, selective pooled FPRs across sample sizes and variance ratios ranged from $4.78\%$ to $5.38\%$, with trial-cluster MCSEs of $0.15$--$0.26$ percentage points. On the same retained full-data targets, naive FPRs ranged from $94.94\%$ to $98.79\%$. Under the four non-null settings, selective coverage at $D=3$, $\lambda=4$ ranged from $94.29\%$ to $95.88\%$, with MCSEs of $0.27$--$0.57$ percentage points. These results support calibration in the stated Gaussian experiment; they are not a simultaneous empirical guarantee for every cell.

The incremental pruning correction is clearest at the stronger penalty. For $n=100$, $\rho=1$, $D=3$ and $\lambda=8$, growth-only inference rejected $14.57\%$ of the 1,270 retained full-data targets, compared with $5.75\%$ for selective inference. Their paired difference was $-8.82$ percentage points (MCSE $1.00$). At $n=800$, the corresponding rates were $8.37\%$ and $4.81\%$ on 6,096 targets, with paired difference $-3.56$ points (MCSE $0.31$). The grown tree, pruned survivor set and targets are identical within each comparison, isolating the consequence of omitting pruning survival. At $\lambda=4$ the additional effect is smaller in some settings; at $\lambda=0$, growth-only and selective calculations agree, as required.

Strong pruning also makes some conditional populations sparse. At $n=100$, $\rho=1$, $D=3$ and $\lambda=8$, Honest inference had 343 computed targets and only 29 endpoint intervals in the fixed subset. Deep fixed-address cells can be smaller still. Such groups remain in the summaries, with their counts and uncertainty, and are not used to claim precise calibration or stable width quantiles.

Pruning does not itself provide tree-wide multiplicity control. For the selective procedure at $n=800$, $\rho=1$, $D=3$, $\lambda=4$, the individual pooled FPR was $4.88\%$, while the probability of at least one rejection was $18.96\%$. Bonferroni reduced the latter to $4.24\%$. All 5,000 trials, including empty trees, enter these tree-level probabilities.

\begin{table}[tbp]
\centering
\caption{Splitwise false-positive rates under a constant treatment effect with known variances, $D=3$ and $\rho=1$. Rates and Monte Carlo standard errors in parentheses are in percent and percentage points, respectively.}\label{tab:dense-calibration}
\begingroup\footnotesize\setlength{\tabcolsep}{3pt}
\begin{tabular}{rrrrrrrr}
\toprule
$n$ & $\lambda$ & $N_{\mathrm{full}}$ & $N_{\mathrm H}$ & Naive & Growth-only & Selective & Honest \\
\midrule
100 & 4 & 7,854 & 2,366 & 95.77 (0.22) & 5.61 (0.28) & 4.95 (0.26) & 4.18 (0.41) \\
200 & 4 & 15,132 & 7,918 & 95.78 (0.16) & 5.70 (0.21) & 5.08 (0.20) & 5.48 (0.26) \\
400 & 4 & 20,129 & 15,099 & 97.68 (0.11) & 5.24 (0.17) & 5.15 (0.17) & 4.95 (0.18) \\
800 & 4 & 23,376 & 19,893 & 98.79 (0.07) & 4.98 (0.16) & 4.88 (0.15) & 5.06 (0.16) \\
100 & 8 & 1,270 & 343 & 100.00 (0.00) & 14.57 (1.00) & 5.75 (0.68) & 3.50 (0.99) \\
200 & 8 & 2,540 & 1,269 & 99.80 (0.09) & 12.20 (0.65) & 5.28 (0.48) & 4.89 (0.61) \\
400 & 8 & 4,315 & 2,608 & 99.93 (0.04) & 9.62 (0.47) & 5.12 (0.38) & 4.45 (0.41) \\
800 & 8 & 6,096 & 4,145 & 99.90 (0.04) & 8.37 (0.36) & 4.81 (0.29) & 4.75 (0.33) \\
\bottomrule
\end{tabular}
\endgroup
\par\smallskip
\parbox{\linewidth}{\footnotesize $N_{\mathrm{full}}$ is the total number of successfully inferred retained splits shared by Naive, Growth-only and Selective; $N_{\mathrm H}$ is the corresponding count for Honest's own trees and holdout targets. Each row uses 5,000 independent trials. Pooled-node standard errors use trials as clusters, including trials without a retained target; the counts are not numbers of independent replications.}
\end{table}

\subsection{Honest comparison: availability and reference-split recovery}\label{sec:sim-honest}
The larger full-data split count describes more opportunities for inference, not necessarily successful detection of heterogeneity. For example, at $\gamma=0$, $n=100$, $\rho=1$, $D=3$, $\lambda=4$, the full-data procedures assessed 7,854 splits and Honest assessed 2,366, over the same 5,000 base trials. The probabilities of any eligible inference were $85.68\%$ and $47.16\%$, respectively. Every target is null in this setting, so availability alone is not a discovery advantage.

\begin{figure}[htbp]
\centering
\includegraphics[width=\linewidth]{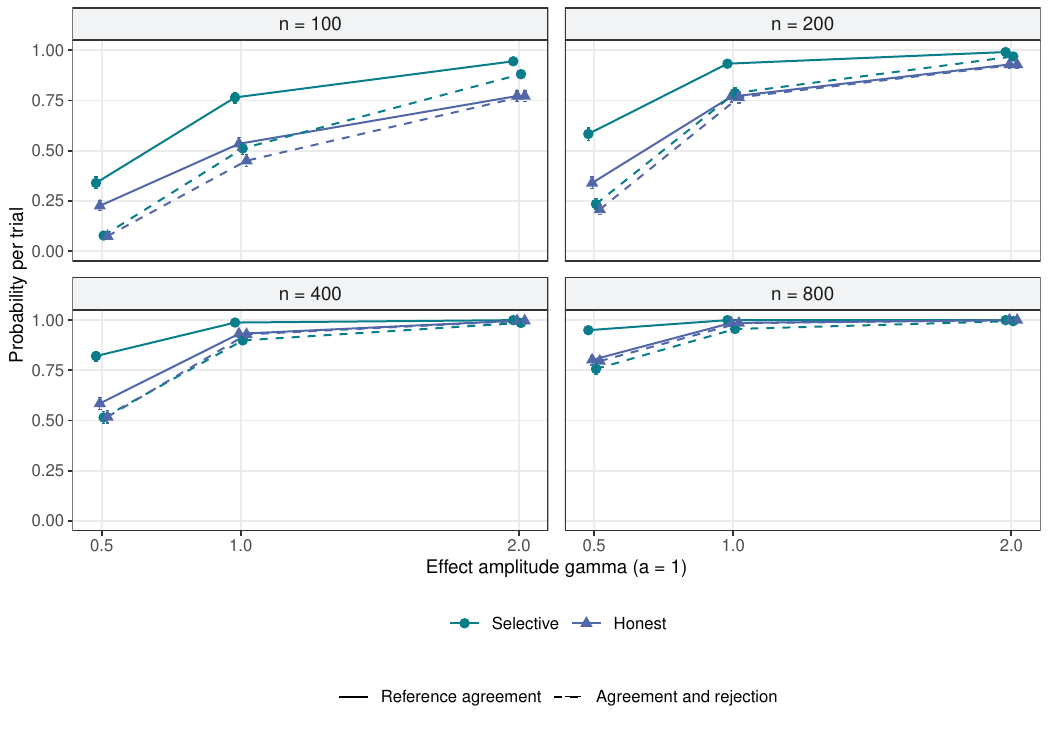}
\caption{Reference-root partition matching and joint recovery, successful inference and rejection, as $\gamma$ varies with $a=1$, $\rho=1$, $D=3$ and $\lambda=4$. Each probability uses all 1,000 trials; bars are pointwise Wilson Monte Carlo intervals. Recovery means ARI greater than $0.75$ under the prespecified common evaluation design, not exact cut-point equality or complete-tree recovery. The two procedures need not infer the same finite-design target.}\label{fig:dense-recovery}
\end{figure}

With genuine heterogeneity, using all outcomes for learning improved reference-partition matching in the small trial. At $(\gamma,a)=(1,1)$, $\rho=1$, $D=3$, $\lambda=4$ and $n=100$, the reference-root recovery rates were $76.5\%$ for the full-data tree and $53.4\%$ for Honest. The joint probabilities of recovery, inference and rejection were $51.1\%$ and $45.0\%$; their MCSEs were $1.58$ and $1.57$ percentage points. These events use all 1,000 trials as denominator. Any eligible inference was available in $100\%$ and $94.7\%$ of these trials, respectively. Figure~\ref{fig:dense-recovery} shows how reference-root matching and joint discovery vary with signal amplitude.

Improved recovery did not imply uniformly improved discovery. At $n=400$ and $800$ in the same setting, recovery-and-rejection probabilities were $89.9\%$ and $95.6\%$ for Selective, versus $92.9\%$ and $98.5\%$ for Honest. Thus learning a reference partition more often and obtaining significant evidence after conditioning are distinct objectives. The procedures' interval widths and general discovery rates also refer to their own selected targets, not a paired comparison of one invariant treatment-effect contrast. Actual third-stage honest splits were absent at $n=100$, as required by the training-size constraint; local ARI matches to deep reference splits do not change this structural fact.

\subsection{Interval widths and their limitations}\label{sec:sim-widths}
Table~\ref{tab:dense-widths} distinguishes actual split stage from maximum growth depth and separates inferential counts from endpoint counts. For $(\gamma,a)=(1,1)$, $\rho=1$, $D=3$ and $\lambda=4$, the root median widths at $n=100$ were $2.216$ for Selective and $2.443$ for Honest, based on 500 and 478 intervals. Their 90th percentiles were $4.225$ and $2.709$. At $n=800$, the root medians were $0.562$ and $0.792$, but the 90th percentiles were $1.087$ and $0.799$. A shorter median therefore coexisted with a longer upper tail.

Deep selective intervals were substantially wider. At $n=400$ in the same setting, third-stage medians were $8.165$ and $2.267$, and 90th percentiles were $47.195$ and $2.924$, based on 1,408 and 754 intervals. These are descriptive comparisons of each procedure's own targets and survivor population. Pooling all stages changes the composition further: additional deep targets can increase the pooled width summary even while root intervals shorten with $n$. Likewise, an alternative-model pooled rejection rate can fall as more weak targets become available; it is not a common-target power curve.

\begin{figure}[htbp]
\centering
\includegraphics[width=\linewidth]{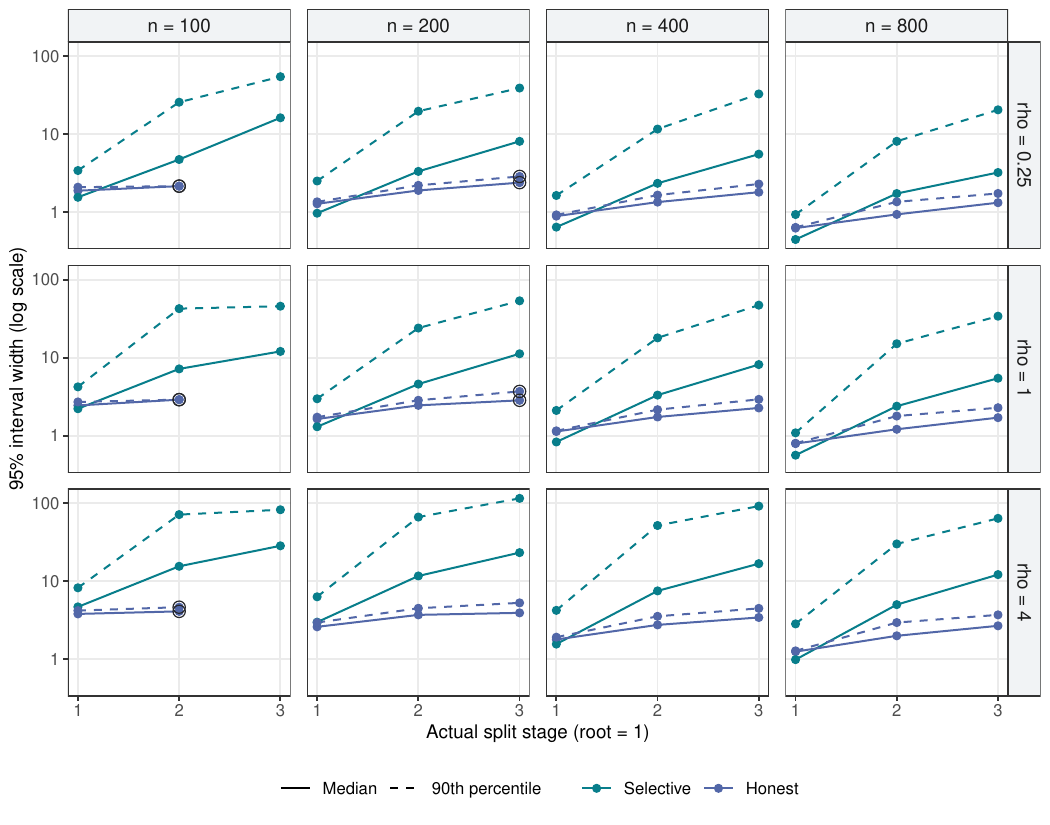}
\caption{Individual 95\% interval widths by actual split stage (root $=1$) for $(\gamma,a)=(1,1)$, $D=3$, $\lambda=4$. Rows specify variance ratio and columns sample size. Curves give medians and 90th percentiles on a logarithmic scale, using all successful endpoints at the fixed indices 1--500 without trimming or capping. Sparse groups with fewer than 30 intervals are flagged; unavailable groups remain missing. These are own-target procedural summaries, not common-parameter efficiency comparisons.}\label{fig:dense-widths}
\end{figure}

Figure~\ref{fig:dense-widths} retains the depth-specific upper-tail behavior. Across the fixed endpoint subsets, the largest selective width with known scale was approximately $3.78\times10^6$, at $n=400$, $\gamma=0$, $\rho=1/4$, $D=3$, $\lambda=4$. The estimated-scale version of the same split had width approximately $3.83\times10^6$. Regenerating the trial, refitting its selection section and checking the endpoints at high precision supported these values; they were not removed or capped. Correct coverage does not ensure useful precision. On the target-score line, ancestor comparisons can impose finite outer boundaries in addition to the central exclusions from growth and pruning (Corollary~\ref{gls:cor:two-components}); this geometry explains a source of inversion sensitivity without diagnosing every long interval or asserting monotonicity in depth.

\begin{table}[tbp]
\centering
\caption{Interval widths by actual split stage with known variances, for $\gamma=a=\rho=1$, $D=3$ and $\lambda=4$. Stage 1 is the root and stage 3 corresponds to node depth two.}\label{tab:dense-widths}
\begingroup\footnotesize\setlength{\tabcolsep}{3pt}
\begin{tabular}{rrlrrrr}
\toprule
$n$ & Stage & Method & $N$ & $N_{\mathrm{CI}}$ & Median & 90th percentile \\
\midrule
100 & 1 & Selective & 1,000 & 500 & 2.22 & 4.23 \\
100 & 1 & Honest & 947 & 478 & 2.44 & 2.71 \\
100 & 3 & Selective & 70 & 40 & 12.04 & 45.73 \\
100 & 3 & Honest & 0 & 0 & --- & --- \\
200 & 1 & Selective & 1,000 & 500 & 1.30 & 2.98 \\
200 & 1 & Honest & 1,000 & 500 & 1.64 & 1.73 \\
200 & 3 & Selective & 1,507 & 750 & 11.23 & 53.65 \\
200 & 3 & Honest & 57 & 25$^{\dagger}$ & 2.84 & 3.70 \\
400 & 1 & Selective & 1,000 & 500 & 0.83 & 2.11 \\
400 & 1 & Honest & 1,000 & 500 & 1.13 & 1.15 \\
400 & 3 & Selective & 2,825 & 1,408 & 8.16 & 47.19 \\
400 & 3 & Honest & 1,507 & 754 & 2.27 & 2.92 \\
800 & 1 & Selective & 1,000 & 500 & 0.56 & 1.09 \\
800 & 1 & Honest & 1,000 & 500 & 0.79 & 0.80 \\
800 & 3 & Selective & 3,363 & 1,679 & 5.46 & 34.07 \\
800 & 3 & Honest & 2,787 & 1,390 & 1.70 & 2.28 \\
\bottomrule
\end{tabular}
\endgroup
\par\smallskip
\parbox{\linewidth}{\footnotesize $N$ counts successfully inferred splits over all 1,000 trials; $N_{\mathrm{CI}}$ counts intervals obtained at the fixed trial indices 1--500. A dagger flags fewer than 30 computed intervals. Widths use all successful intervals in that subset without capping or trimming. The two procedures generally have different projection targets and survivor populations, including within the same stage. The Honest training sample at $n=100$ cannot form a third-stage split under the common minimum-cell rule; its unavailable quantiles are shown as dashes.}
\end{table}

\subsection{Substituting an estimated common scale}\label{sec:sim-plugin}
The additional module keeps the variance ratio $\rho$ known and estimates only the common scale. Regressing control outcomes on the prespecified basis gives
\begin{equation}\label{eq:sim-sigmahat}
\widehat\sigma^2=\frac{\RSS_0}{n_0-\operatorname{rank}(B_0)},
\end{equation}
where $B_0$ is the control-group basis matrix. For the full-data procedures this uses all controls; Honest uses training controls only and keeps the estimate fixed during holdout inference. Control outcomes contain no treatment-effect heterogeneity, and their prognosis is correctly specified here. The estimate need not be independent of the learned tree. We substitute $\widehat\sigma$ only in the truncated-normal reference distribution, leaving $V$, the raw numerical $\lambda$, the tree and the target unchanged. The comparison covers $\gamma=0$ and $(\gamma,a)=(1,1)$, all sample sizes, ratios and depth limits, at $\lambda=4$. It does not estimate both arm variances or establish exactness with unknown covariance.

\begin{figure}[htbp]
\centering
\includegraphics[width=\linewidth]{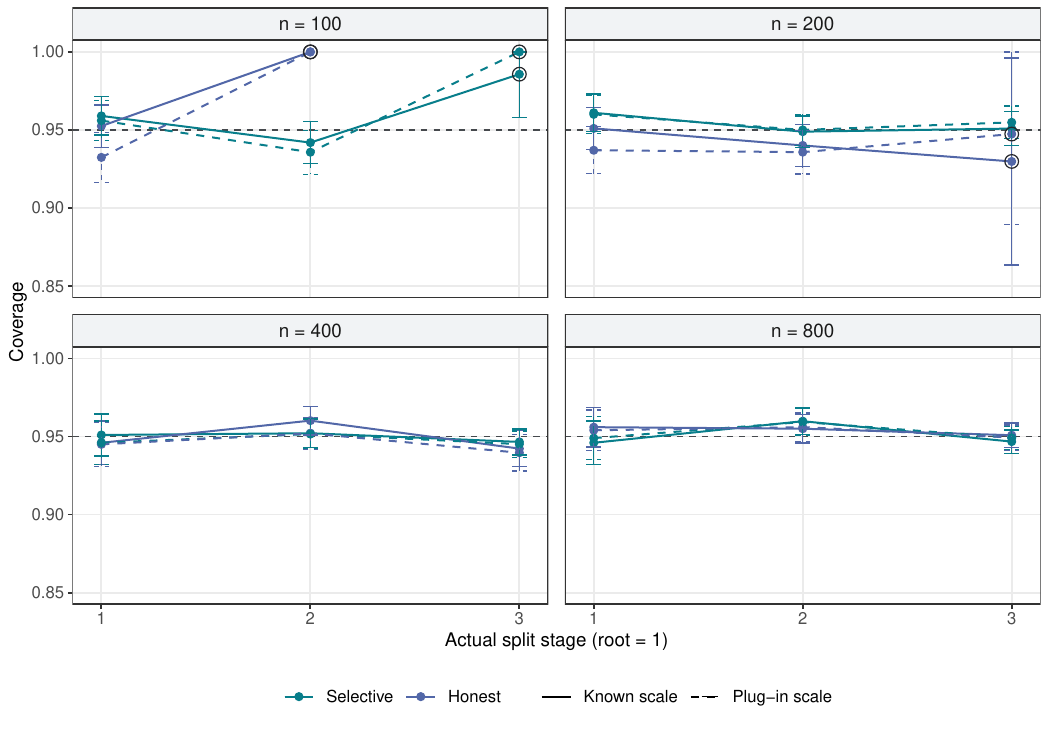}
\caption{Known and estimated common-scale coverage by actual split stage for $(\gamma,a)=(1,1)$, $\rho=1$, $D=3$ and $\lambda=4$. Panels specify sample size. Known/estimated comparisons use identical trees, targets and eligibility within a procedure; Honest has its own holdout targets. Bars are pointwise trial-cluster Monte Carlo intervals; rings mark fewer than 100 computed targets or 30 contributing trials. The dashed reference is $0.95$. The displayed vertical range is a viewing window, not a truncation of the computed rates; full estimates and intervals remain in the reproducibility summaries. No exactness claim is made for scale substitution.}\label{fig:dense-plugin}
\end{figure}

For Selective at $D=3$, the largest reduction in coverage relative to known scale was $0.36$ percentage points under the constant effect and $0.46$ points under $(\gamma,a)=(1,1)$. Table~\ref{tab:dense-plugin} reports method-specific counts and paired changes, and Figure~\ref{fig:dense-plugin} separates actual split stages. At $n=100$, $\rho=1$, $(\gamma,a)=(1,1)$, Selective coverage changed from $95.10\%$ to $94.70\%$ on the same 2,206 targets. Honest coverage changed from $95.26\%$ to $93.26\%$ on its own 949 targets, a paired difference of $-2.00$ percentage points (MCSE $0.56$). At $\rho=4$, the honest paired change was $-2.50$ points (MCSE $0.61$) on 801 targets. These results indicate limited sensitivity for the specified Selective scale estimator in this experiment and a detectable small-sample effect of scale substitution for Honest. They remain empirical observations under known variance shape and correct prognosis, separate from the finite-sample known-covariance guarantee.

\begin{table}[tbp]
\centering
\caption{Coverage after substituting the estimated common scale, for $\gamma=a=1$, $D=3$ and $\lambda=4$.}\label{tab:dense-plugin}
\begingroup\footnotesize\setlength{\tabcolsep}{3pt}
\begin{tabular}{rrlrrrr}
\toprule
$n$ & $\rho$ & Method & $N$ & Known scale & Estimated scale & Difference \\
\midrule
100 & $1/4$ & Selective & 2,353 & 95.28 (0.45) & 94.86 (0.47) & -0.42 (0.22) \\
100 & $1/4$ & Honest & 988 & 95.04 (0.69) & 93.22 (0.80) & -1.82 (0.57) \\
100 & 1 & Selective & 2,206 & 95.10 (0.47) & 94.70 (0.48) & -0.41 (0.26) \\
100 & 1 & Honest & 949 & 95.26 (0.69) & 93.26 (0.81) & -2.00 (0.56) \\
100 & 4 & Selective & 1,956 & 95.35 (0.52) & 94.89 (0.53) & -0.46 (0.31) \\
100 & 4 & Honest & 801 & 96.38 (0.66) & 93.88 (0.85) & -2.50 (0.61) \\
800 & $1/4$ & Selective & 6,346 & 95.23 (0.28) & 95.18 (0.28) & -0.05 (0.08) \\
800 & $1/4$ & Honest & 5,906 & 95.41 (0.28) & 95.46 (0.27) & +0.05 (0.12) \\
800 & 1 & Selective & 6,354 & 95.07 (0.28) & 95.23 (0.28) & +0.16 (0.08) \\
800 & 1 & Honest & 5,743 & 95.32 (0.27) & 95.25 (0.27) & -0.07 (0.14) \\
800 & 4 & Selective & 6,138 & 95.00 (0.29) & 94.90 (0.30) & -0.10 (0.09) \\
800 & 4 & Honest & 5,343 & 94.95 (0.30) & 94.78 (0.31) & -0.17 (0.13) \\
\bottomrule
\end{tabular}
\endgroup
\par\smallskip
\parbox{\linewidth}{\footnotesize Coverage is in percent; parentheses give trial-cluster Monte Carlo standard errors in percentage points. Difference is estimated minus known coverage, with its paired trial-cluster standard error. Within each method the two scale versions use the same data, trees, targets and $N$ inferred splits. Honest is evaluated on its own holdout targets and estimates the scale using training controls only. These plug-in calculations retain the known variance ratio and do not carry a finite-sample exactness guarantee.}
\end{table}

\FloatBarrier

\section{Real-data application: ACTG 175}\label{sec:actg}

\subsection{Scientific question and public analysis sample}
ACTG 175 compared four nucleoside regimens in adults with HIV infection \citep{Hammer1996}. We examine the historical comparison of zidovudine (ZDV) plus didanosine (ddI) with ZDV plus zalcitabine (zal, also denoted ddC). The scientific question is whether combinations of baseline immune status, age, and prior-treatment history describe an adjusted interaction in early immunologic response. We use the CD4 count at $20\pm5$ weeks on its original scale, in cells per cubic millimeter. This quantitative response differs from the original trial's primary endpoint combining CD4 decline, AIDS progression, and death. The companion biomarker investigation supplies the context for early immune response \citep{Katzenstein1996}. The same two combination regimens and 20-week response were used by \citet[Section 4]{Fan2017}; our search rule, nuisance basis, and projection targets are those of the present method.

The public \texttt{ACTG175} file in \texttt{speff2trial} version 1.0.5 contains 2139 participants and 27 variables \citep{speff2trial2022}. Restricting \texttt{arms} to 1 and 2 gives 522 ZDV+ddI and 524 ZDV+zal assignments, respectively, for $n=1046$. We set $A=1$ for ZDV+ddI and $A=0$ for ZDV+zal. All selected baseline variables and 20-week responses are recorded in this public subset. The original trial reported 2467 participants; \citet[Section 5.2]{Huang2020} describe the 2139 records as remaining after exclusion of missing or unrecorded relevant information, without identifying the complete exclusion rule. We therefore do not interpret public-file completeness as complete original-trial follow-up or claim to reproduce the original intention-to-treat analysis. Three records have baseline CD4 equal to zero and are retained as supplied; the inspected dictionary does not identify zero as a missing-value code. No response transformation, outcome imputation, or outlier deletion is used.

\subsection{Maximum-depth-three unpruned analysis}
The main display is an \emph{unpruned} interaction tree with maximum depth three, root depth zero, and dimensionless penalty $\widetilde\lambda=0$. We report every eligible grown internal split. Maximum depth is an upper bound: a region stops earlier when its design has no eligible candidate. With zero penalty, growth-only and pruning-aware calculations coincide for the observed trees, so the main comparison has three distinct procedures: Naive, Selective, and Honest. Here Selective denotes the common growth-adjusted and pruning-aware result, rather than two separate empirical competitors. The additional contribution of correcting positive-penalty pruning is examined in the theoretical results and controlled simulations; positive-penalty application calculations are summarized below as exploratory references.

Earlier exploration of these public data had already included outcome-distribution diagnostics, restricted-grid trees and several fixed penalties. The present expansion to all adjacent-observation cut points was adopted after those results had been inspected. Before this rerun, we retained the original-scale response, common basis, maximum depth, minimum-cell rule, saved Honest partition and unpruned main display, and fixed the expanded candidate policy and all four positive-penalty references. These choices are not an outcome-unseen analysis plan. Neither candidate-policy selection nor selection of the response scale or main display is included in the growth-conditioning event, and no additional adjustment for these exploratory choices is claimed.

The common basis contains an intercept, baseline CD4, baseline CD8, age, weight, Karnofsky score, recorded sex, symptom indicator, prior antiretroviral days, and two indicators for the three-level prior-treatment stratum. Nonconstant columns are centered and scaled using the 1046 baseline records, preserving the column space. Every candidate uses $C_s=[B_R,A_R,h_s]$ as in~\eqref{pr:eq:contrast}; rank deficiency is allowed, but the interaction coefficient must be identifiable. The six search variables, in fixed priority order, are baseline CD4, baseline CD8, age, prior-treatment days, prior experience, and symptoms. Within each reached parent, every midpoint between distinct adjacent observed values of a continuous search variable is considered; the binary variables have threshold $0.5$. The full-data tree constructs candidates from its parent observations, whereas Honest constructs its thresholds from training covariates alone and applies those same numerical thresholds to holdout participants. Each treatment-by-child cell must contain at least 20 participants during growth. Exact duplicate or complementary child partitions use their first fixed-priority representative. Post-randomization variables are excluded.

The variances are unknown. Groupwise regressions on the fixed basis give $s_a^2=\RSS_a/\{n_a-\operatorname{rank}(B_a)\}$. The full-data calculations freeze $\widehat V_{ii}=s_{A_i}^2$ and $\sigma=1$ for growth and inference, with estimated standard deviations $104.8919$ and $132.0497$ for ZDV+zal and ZDV+ddI, respectively. Variances are not re-estimated within selected leaves. Honest uses one saved 523/523 split, generated from sorted participant IDs with seed 20261028: training arm counts are 263/260 and holdout counts are 261/262, ordered as ZDV+zal/ZDV+ddI. Its two variance pilots use training outcomes alone, giving standard deviations $101.2552$ and $146.5730$, and are frozen for both training growth and holdout inference. A held-out contrast requires at least 15 observations in its parent, at least one in each treatment-by-child cell, and an identifiable interaction coefficient. Figure~\ref{fig:actg-tree} reports the complete split rules and arm counts.

\subsection{Unpruned results and positive-penalty reference analyses}
\begin{figure}[htbp]\centering
\includegraphics[width=0.98\linewidth]{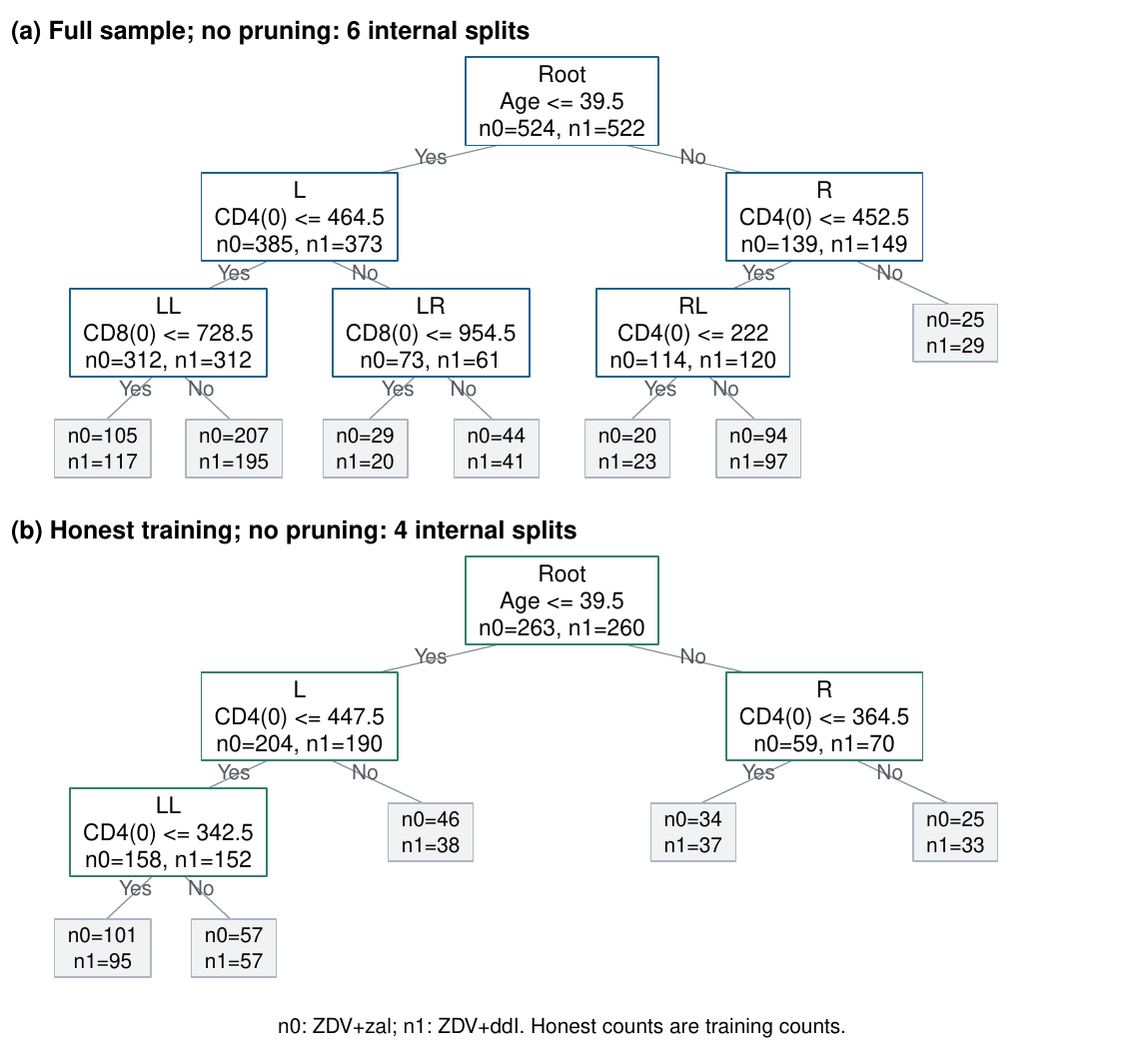}
\caption{ACTG 175 all-midpoint unpruned trees at $\lambda=0$ and maximum depth three. Full-data growth gives 6 internal splits; Honest training gives 4. Labels show the learned rules and arm counts, with $n_0$ for ZDV+zal and $n_1$ for ZDV+ddI. Honest counts are training counts. Each part generates its candidate thresholds from its own covariates. Addresses label each tree's own regions and do not identify shared targets across trees.}\label{fig:actg-tree}
\end{figure}

At $\lambda=0$, all 6 full-data grown splits and all 4 Honest training splits remain. Their distinct eligible root-candidate counts are 1,443 and 828, respectively. Of the training-selected Honest splits, 4 support inference in their holdout regions. Figure~\ref{fig:actg-tree} gives both unpruned trees, and Table~\ref{tab:actg-main} reports every primary-display record: 6 Naive, 6 Selective and 4 Honest. Growth-only and pruning-aware calculations have identical probabilities, intervals and $p$-values at zero penalty and are displayed once as Selective.

The full-data root rule is $\mathrm{Age}\leq 39.5$, with GLS interaction estimate $-65.50$. Its naive nominal 95\% interval is $[-97.97,-33.04]$ ($p<0.001$), compared with the Selective interval $[-92.12,21.67]$ ($p=0.651$). Across the full-data tree, 5 of 6 Naive intervals and 1 of 6 Selective intervals exclude zero. These changes concern nominal evidence after the observed search, not demonstrated false positives or known clinical truth.
The sole Selective exclusion is at node R, with rule $\mathrm{CD4}_0\leq 452.5$ in that node's parent, estimate $-103.50$ and interval $[-174.06,-8.05]$ ($p=0.036$).

The Honest root rule is $\mathrm{Age}\leq 39.5$, geometrically agreeing with the full-data root. Its holdout-design estimate is $-38.87$, with interval $[-86.25,8.52]$ ($p=0.108$). One of the 4 available Honest intervals excludes zero. The Honest exclusion occurs at node LL, with rule $\mathrm{CD4}_0\leq 342.5$, holdout estimate $-69.40$ and interval $[-131.32,-7.49]$ ($p=0.028$). This is a different rule and target from the full-data nominal exclusion.

\begin{figure}[htbp]\centering
\includegraphics[width=0.98\linewidth]{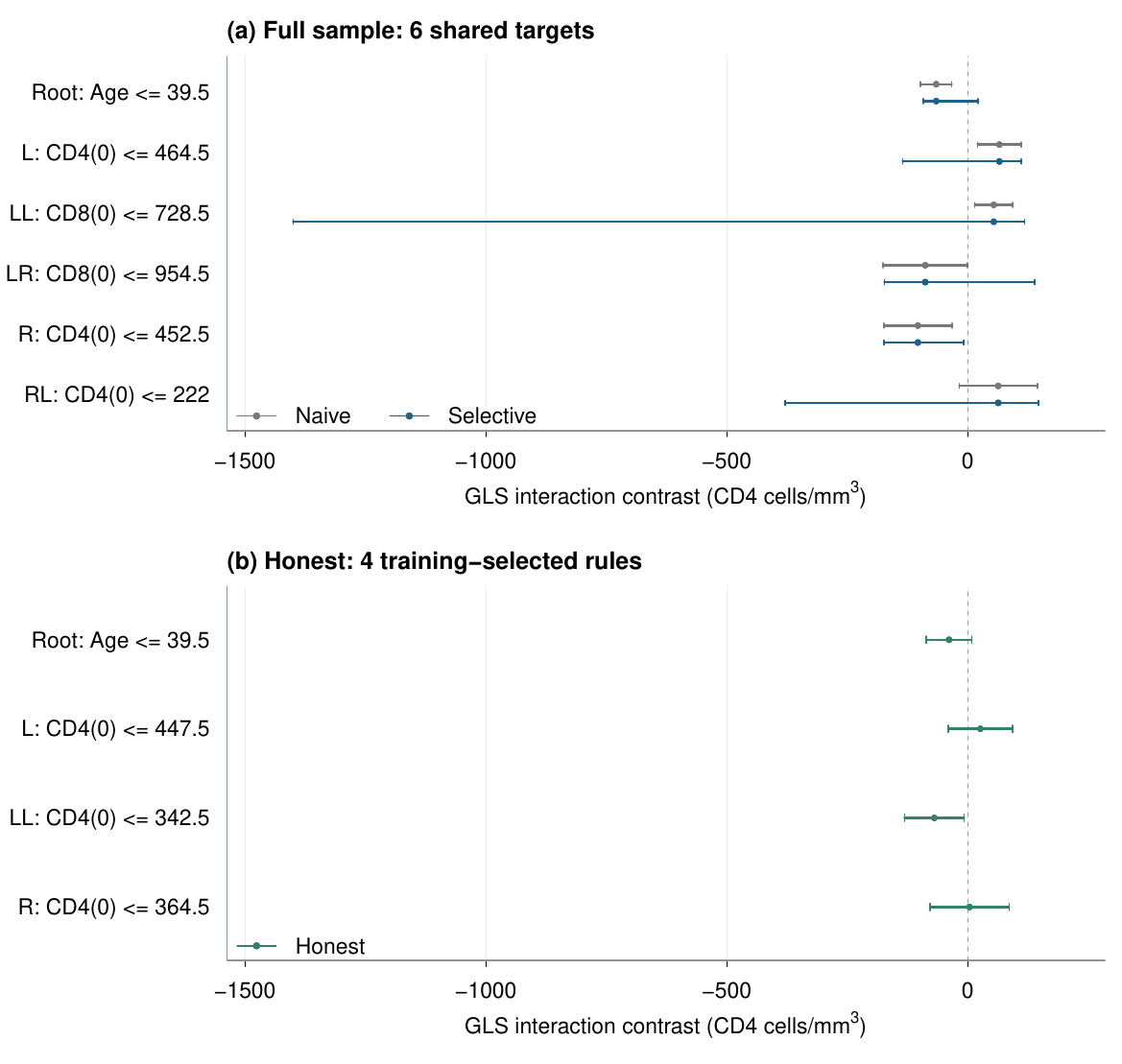}
\caption{All 16 primary-display nominal individual 95\% intervals from the full-midpoint analysis. Naive and Selective share the 6 full-data targets; their zero-penalty Growth-only duplicates are not drawn separately. Honest has 4 available holdout-design targets learned from training data. Both panels use the same horizontal scale. Dots are unadjusted GLS point estimates, and every finite confidence endpoint is shown. The displayed working-model intervals with estimated covariance are neither established exact nor simultaneous clinical intervals.}\label{fig:actg-intervals}
\end{figure}

The full-data Selective and Honest median widths are $279.38$ and $128.30$, and their maxima are $1518.14$ and $164.23$, respectively. The rules, working weights and analysis designs can differ, so these are descriptive procedural comparisons rather than efficiencies for a common clinical parameter. Figure~\ref{fig:actg-intervals} displays every endpoint without clipping.

For fixed positive penalties $\lambda=1,2,3,4$, the numbers of retained full-data splits are 6, 6, 5, 4, and the corresponding Honest training counts are 4, 4, 4, 3. At $\lambda=3$ and $4$, the retained full-data R contrast keeps its estimate of $-103.50$, but its pruning-aware $p$-values become $0.098$ and $0.180$; growth-only inference remains at $p=0.036$. Thus no Selective interval excludes zero at these two penalties. Honest retains its LL exclusion through penalty three, whereas penalty four removes that split. All reference results are retained in the reproducibility files. Positive penalties may alter the truncation section even when the retained tree is unchanged. The largest Selective reference width is $1518.14$. These are exploratory reference calculations; no penalty is selected from its nominal $p$-value.

\begin{table}[tbp]\centering\footnotesize
\caption{ACTG 175 full-midpoint unpruned display at maximum depth three and $\lambda=0$: 6 full-data splits and 4 Honest training splits. All intervals and $p$-values are individual nominal working-model calculations. Growth-only and Selective coincide at zero penalty and are displayed once as Selective. CD4 units are cells per cubic millimeter.}\label{tab:actg-main}
\setlength{\tabcolsep}{3pt}
\begin{tabular}{lllrlrr}\toprule
Node & Rule & Method & Estimate & 95\% interval & $p$ & Width \\
\midrule
Root & $\mathrm{Age}\leq 39.5$ & Naive & -65.50 & [-97.97, -33.04] & $<0.001$ & 64.93 \\
 &  & Selective & -65.50 & [-92.12, 21.67] & $0.651$ & 113.79 \\
\addlinespace[2pt]
L & $\mathrm{CD4}_0\leq 464.5$ & Naive & 65.68 & [20.28, 111.09] & $0.005$ & 90.81 \\
 &  & Selective & 65.68 & [-135.21, 111.15] & $0.566$ & 246.37 \\
\addlinespace[2pt]
LL & $\mathrm{CD8}_0\leq 728.5$ & Naive & 54.00 & [14.64, 93.36] & $0.007$ & 78.72 \\
 &  & Selective & 54.00 & [-1400.38, 117.76] & $0.315$ & 1518.14 \\
\addlinespace[2pt]
LR & $\mathrm{CD8}_0\leq 954.5$ & Naive & -88.20 & [-176.03, -0.36] & $0.049$ & 175.66 \\
 &  & Selective & -88.20 & [-173.20, 139.18] & $0.437$ & 312.38 \\
\addlinespace[2pt]
R & $\mathrm{CD4}_0\leq 452.5$ & Naive & -103.50 & [-174.14, -32.85] & $0.004$ & 141.30 \\
 &  & Selective & -103.50 & [-174.06, -8.05] & $0.036$ & 166.01 \\
\addlinespace[2pt]
RL & $\mathrm{CD4}_0\leq 222$ & Naive & 63.21 & [-17.88, 144.30] & $0.127$ & 162.18 \\
 &  & Selective & 63.21 & [-379.40, 146.71] & $0.937$ & 526.11 \\
\addlinespace[2pt]
\midrule
Root & $\mathrm{Age}\leq 39.5$ & Honest & -38.87 & [-86.25, 8.52] & $0.108$ & 94.77 \\
L & $\mathrm{CD4}_0\leq 447.5$ & Honest & 26.16 & [-40.23, 92.56] & $0.440$ & 132.78 \\
LL & $\mathrm{CD4}_0\leq 342.5$ & Honest & -69.40 & [-131.32, -7.49] & $0.028$ & 123.83 \\
R & $\mathrm{CD4}_0\leq 364.5$ & Honest & 3.75 & [-78.37, 85.86] & $0.929$ & 164.23 \\
\bottomrule\end{tabular}
\par\smallskip\parbox{\linewidth}{\footnotesize Addresses identify each tree's own nodes. The full-data Naive and Selective results share targets, whereas Honest uses its own training-selected rules and holdout-design targets. Unavailable holdout contrasts are distinguished from numerical failure. Intervals are not simultaneous, and estimated variances do not establish exact coverage.}
\end{table}

\subsection{Interpretation and limits}
All application intervals and $p$-values are \emph{nominal working-model calculations}. Treating estimated group variances as fixed defines the computation but does not satisfy Assumption~\ref{pr:ass:gaussian}. Raw CD4 is a nonnegative count response. Groupwise residual diagnostics show right-tail departures, particularly in the ZDV+zal reference group; mean-model lack of fit and treatment-effect variation can also contribute to these residual patterns. Neither the raw-scale choice nor the sample size establishes a conditional Gaussian law. Training-only variance estimation prevents holdout leakage into selection but does not supply a known-true-variance guarantee for Honest. Algebraic agreement of compatible path and full-tree sections is not evidence of exact clinical coverage under these departures.

The target $\eta_s\trans\mu$ is a finite-design GLS interaction specific to its selected split, basis, allocation, and working variance ratio. Its weighted-CATE interpretation additionally requires causal identification in this analyzed subset and a correct prognosis under the new reference regimen, ZDV+zal. The unknown public-file inclusion rule prevents us from deriving these conditions from original randomization alone. Facility identifiers are absent, limiting assessment of residual dependence. Full-data and Honest estimates can concern different rules and holdout-design targets, so their widths are not paired efficiencies for one invariant clinical parameter. All displayed intervals are individual 95\% intervals, with no demonstrated tree-wide familywise guarantee. This single observed dataset estimates neither coverage nor false-positive rates. The exploratory comparisons and nominal exclusions of zero do not establish treatment recommendations.
\FloatBarrier

\section{Discussion}\label{sec:discussion}
We developed selective inference for pruned, covariate-adjusted interaction trees with known unequal outcome variances. Building on the Tree-Values framework \citep{Neufeld2022}, we established common-basis GLS orthogonality for residualized treatment interactions. Under additive fixed-penalty pruning, this geometry reduces retention of a selected path to its growth constraints and one quadratic pruning-survival inequality, yielding exact finite-sample conditional tests and confidence intervals under the stated Gaussian model. We also showed that retained-path and complete-tree conditioning give identical inference on compatible sections. This factorization simplifies computation while preserving the complete-tree-conditioned intervals.

The theory identifies a finite-design GLS target that remains unchanged when a split survives pruning. Under causal identification and a correct reference prognosis, this target is a signed weighted contrast of conditional average treatment effects. Its weights depend on allocation, covariates and variance shape, so it generally differs from an ordinary subgroup-average treatment-effect contrast. This distinction connects the inferential guarantee to the specific interaction being estimated and clarifies the additional assumptions needed for a causal interpretation.

Simulations at $n=100,200,400,800$, with maximum depths one to three and all adjacent-observation candidate midpoints, demonstrated the practical importance of correcting both growth and pruning. Known-variance results supported nominal calibration. For example, at $n=100$, $\rho=1$, $D=3$ and $\lambda=8$ under a constant effect, growth-only inference rejected $14.57\%$ of retained contrasts, compared with $5.75\%$ after pruning correction on the same targets (Section~\ref{sec:sim-calibration}). Full-data learning improved reference-split recovery in some small-sample settings, although subsequent discovery probabilities were not uniformly higher. Root-level median intervals could be shorter than Honest intervals, while deeper intervals and upper tails were substantially wider. These comparisons describe each procedure's selected targets and populations.

The Honest comparison highlights the value of preserving data for exploration in small samples. Dividing observations between training and inference reduces the information available to each task, and local minimum-cell requirements can prevent further refinement. In ACTG 175, requiring 20 observations in each treatment-by-child cell requires at least 40 in each treatment arm of a parent. The Honest training region LR contained 46 and 38 observations and could not be split further. The full-data and Honest trees had six and four internal splits, respectively (Figure~\ref{fig:actg-tree}). Full-data exploration can therefore preserve opportunities for finer partitioning, while the wide nominal Selective intervals in Figure~\ref{fig:actg-intervals} show the limits to the precision of the resulting inference.

Honest results also depend on the particular training/holdout assignment, which can change the eligible candidates, learned groups, inferential targets and intervals. This dependence can be a practical concern for analysts. A fixed seed makes one realization reproducible, but does not establish stability across alternative partitions; we did not quantify this sensitivity. The full-data procedure avoids this additional random partition, although its tree can still be sensitive to changes in the observed sample. The combination of richer exploration and selection adjustment is thus a practical advantage, with useful precision and stable subgroup discovery remaining separate objectives.

The exact guarantees require a conditional Gaussian model with known positive diagonal covariance and prespecified growth and pruning rules. They concern individual selected contrasts; simultaneous reporting requires an additional multiplicity adjustment. Our estimated-scale experiment retained a known variance ratio and showed modest changes in Selective coverage in the settings examined, without establishing unknown-variance validity. ACTG 175 uses estimated arm variances and has residual tail departures, unresolved public-subset inclusion and prior exploratory analysis choices. Its results remain nominal working-model illustrations. Developing valid theory for unknown arm-specific variances, and assessing robustness to prognosis misspecification and non-Gaussian outcomes, are central practical extensions.

Further work could improve how information is allocated between discovery and inference. Data thinning can construct independent components of each response while retaining the covariate design \citep{Neufeld2024Thinning}; data carving can use information remaining in the selection sample after conditioning \citep{Fithian2014}; and randomized selection offers another way to retain information for subsequent inference \citep{Tian2018}. Incorporating these strategies into interaction-tree growth and pruning may enable more efficient estimation and inference. Their adaptation requires new selection laws and evaluation of estimation error, power and interval precision. Gaussian thinning itself relies on known covariance, so it does not resolve the unknown-variance problem. Establishing when these extensions improve performance in small samples is an important direction for future research.

\FloatBarrier
\section*{Acknowledgement}\label{sec:acknowledgement}

\def\coderepositoryurl{https://github.com/tomoshige/selective-inference-for-interaction-tree}

\paragraph{Data and computation.}
The ACTG 175 records were obtained from the publicly distributed \texttt{speff2trial} package, version 1.0.5 \citep{speff2trial2022}. The application is a secondary analysis of those public records; no participants were recruited or contacted. The simulations use generated outcomes and do not involve human participants. Appendices~\ref{supp:sec:proofs} and~\ref{supp:sec:extensions} contain the complete proofs and further structural results.

\paragraph{Code and reproducibility.}
The accompanying reproducibility package provides the simulation data-generation and inference code, the public ACTG 175 data, and English instructions for running both workflows and generating the reported tables and figures. Results are generated when the code is run.
\ifx\coderepositoryurl\empty
\else
These materials are available in the \href{\coderepositoryurl}{GitHub repository}.
\fi

\paragraph{Use of generative artificial intelligence.}
OpenAI Codex assisted with manuscript drafting, analytical derivations, simulation and data analysis programming, numerical verification, and preparation of figures and tables. The authors are responsible for checking and approving all scientific content, code, references, and interpretations before submission.

\clearpage
\appendix
\counterwithin{equation}{section}
\counterwithin{table}{section}
\counterwithin{figure}{section}
\counterwithin{theorem}{section}
\counterwithin{proposition}{section}
\counterwithin{lemma}{section}
\counterwithin{corollary}{section}
\counterwithin{definition}{section}
\counterwithin{remark}{section}
\section{Proofs of the main results}\label{supp:sec:proofs}

\subsection{Proof of Proposition~\ref{pr:prop:optimizer}}
\begin{proof}
At a terminal node the only admissible score is zero. At an internal node, collapsing gives zero; retaining gives its penalized gain plus the optimal scores in its two children. The maximum is precisely~\eqref{pr:eq:dp}. If $V_w<0$, every optimum collapses. If $V_w=0$, collapsing has zero internal nodes, fewer than any retained alternative. If $V_w>0$, every optimum retains $w$ and uses optimal subtrees in both children. By induction, the minimum-node choices in the children are unique, proving both optimality and uniqueness. One bottom-up pass computes the values and one top-down pass reconstructs the tree.
\end{proof}

\subsection{Proof of Proposition~\ref{pr:prop:causal}}
\begin{proof}
Fix the design and an eligible candidate $s$ in its parent region $R$, so that $d_s>0$. We first work with vectors restricted to $I_R$. Write $\eta_{s,R}=W_Rv_s/d_s^2$ for the local contrast; the full vector $\eta_s$ is its zero extension outside $I_R$.

\emph{Step 1: orthogonality and normalization of the contrast.}
Set $\widetilde C_s=V_R^{-1/2}C_s$ and $\widetilde g_s=V_R^{-1/2}g_s$. By definition,
\[
r_s=(I-P_{\widetilde C_s})\widetilde g_s.
\]
An orthogonal projection leaves a residual orthogonal to its column space, hence $\widetilde C_s\trans r_s=0$. Since $v_s=V_R^{1/2}r_s$ and $W_R=V_R^{-1}$, this gives the original-coordinate identity
\[
C_s\trans W_Rv_s
=C_s\trans V_R^{-1/2}r_s
=\widetilde C_s\trans r_s=0.
\]
The fitted vector $P_{\widetilde C_s}\widetilde g_s$ belongs to $\col(\widetilde C_s)$, so it equals $\widetilde C_sa_s$ for some $a_s$. Such coefficients exist even when $C_s$ is rank deficient; their uniqueness is not required. Multiplying the residual decomposition by $V_R^{1/2}$ yields $g_s=v_s+C_sa_s$. Consequently,
\[
v_s\trans W_Rg_s
=v_s\trans W_R(v_s+C_sa_s)
=v_s\trans W_Rv_s=d_s^2.
\]
Dividing by $d_s^2>0$ establishes
\[
\eta_{s,R}\trans C_s=0,
\qquad \eta_{s,R}\trans g_s=1.
\]
Because $C_s=[B_R,A_R,h_s]$ and $B_R$ contains an intercept, we therefore have $\eta_{s,R}\trans B_R=0$, $\eta_{s,R}\trans A_R=0$, $\eta_{s,R}\trans h_s=0$, and $\eta_{s,R}\trans\ind_{I_R}=0$, where $\ind_{I_R}$ is the all-ones vector on the parent.

\emph{Step 2: the four treatment-by-child cell sums.}
Recall that $h_s=1$ in the left child and $h_s=0$ in the right child. For $a\in\{0,1\}$, define
\[
\mathcal I_{aL}=\{i\in I_R:A_i=a,\ h_{s,i}=1\},
\qquad
\mathcal I_{aR}=\{i\in I_R:A_i=a,\ h_{s,i}=0\},
\]
and let $w_{aC}=\sum_{i\in\mathcal I_{aC}}\eta_{s,i}$ for $C\in\{L,R\}$. The four identities from Step 1 determine these sums successively. In particular, $g_s=A_R\odot h_s$ is the indicator of the left-treated cell, so
\[
\begin{aligned}
w_{1L}&=\eta_{s,R}\trans g_s=1,\\
w_{0L}&=\eta_{s,R}\trans h_s-w_{1L}=0-1=-1,\\
w_{1R}&=\eta_{s,R}\trans A_R-w_{1L}=0-1=-1,\\
w_{0R}&=\eta_{s,R}\trans\ind_{I_R}-w_{1L}-w_{0L}-w_{1R}
=0-1+1+1=1.
\end{aligned}
\]
These are identities for cell totals; they do not impose a common sign on all individual weights within a cell.

\emph{Step 3: causal interpretation and within-child homogeneous effects.}
Zero extension gives $\eta_s\trans B=\eta_{s,R}\trans B_R=0$. Under Assumption~\ref{pr:ass:causal}, the mean decomposition in~\eqref{pr:eq:causal-mean} therefore implies
\[
\begin{aligned}
\theta_s
&=\eta_s\trans\mu
=\eta_s\trans B\beta+\eta_s\trans(A\odot\tau_X)\\
&=\sum_{i\in I_R:A_i=1}\eta_{s,i}\tau(X_i).
\end{aligned}
\]
This is~\eqref{pr:eq:weighted-target}. Now suppose $\tau(X_i)=\tau_L$ at every realized design point in the left child and $\tau(X_i)=\tau_R$ at every realized design point in the right child, as stipulated in the proposition. The treated-cell sums from Step 2 give
\[
\theta_s=\tau_Lw_{1L}+\tau_Rw_{1R}
=\tau_L-\tau_R.
\]
If the effect is constant throughout the parent, then $\tau_L=\tau_R$, and hence $\theta_s=0$.

\emph{Step 4: invariance under pruning.}
Consider pruning that retains $s$ with its original parent and immediate child partition. It leaves $I_R$, $B_R$, $A_R$, $h_s$, and $V_R$ unchanged. Thus $C_s$, $g_s$, their unique projection residual, and $d_s$ are unchanged, even if the nuisance coefficients are not unique. The original contrast $\eta_s$ and its target $\eta_s\trans\mu$ are consequently unchanged. This argument concerns the original local contrast, rather than a new effect estimate obtained by refitting the final leaves.

The orthogonality, normalization, and pruning arguments are deterministic. Assumption~\ref{pr:ass:causal} is used for the causal mean representation in Step 3; Gaussian sampling is not needed for this proposition.
\end{proof}

\subsection{Proof of Lemma~\ref{pr:lem:orthogonality}}
\begin{proof}
Fix the design and the candidate labels. For a candidate $q$ in a parent region $Q$, write $v_{q,Q}$ for its residual restricted to $I_Q$ and $v_q$ for its zero extension to all $n$ observations. We use analogous restriction subscripts below. Only eligible candidates are considered, so $d_q>0$. We first prove the deterministic identities and then derive their sampling consequences.

\emph{Step 1: local weighted normal equations.}
For any such candidate, set $\widetilde C_q=V_Q^{-1/2}C_q$ and $\widetilde g_q=V_Q^{-1/2}g_q$. The construction in~\eqref{pr:eq:contrast} gives
\[
v_{q,Q}=V_Q^{1/2}(I-P_{\widetilde C_q})\widetilde g_q.
\]
The projection residual is orthogonal to every column of $\widetilde C_q$. Because $W_Q=V_Q^{-1}$, this implies
\[
C_q\trans W_Qv_{q,Q}
=\widetilde C_q\trans(I-P_{\widetilde C_q})\widetilde g_q=0.
\]
Moreover, $P_{\widetilde C_q}\widetilde g_q$ belongs to $\col(\widetilde C_q)$, so there exists a coefficient vector $a_q$ such that
\[
v_{q,Q}=g_q-C_qa_q.
\]
The residual is unique and these identities hold even if $C_q$ is rank deficient; uniqueness of $a_q$ is unnecessary. Since $C_q=[B_Q,A_Q,h_q]$, the residual is, in particular, weighted-orthogonal to every column of $[B_Q,A_Q]$.

\emph{Step 2: orthogonality for nested regions.}
Consider the split $s$ in $R$ and any region $D'$ contained in one of its children. On $D'$, the indicator $h_s$ is a constant $e\in\{0,1\}$, and $g_s=A_R\odot h_s$ restricts to $eA_{D'}$. Partition the coefficients from Step 1 as $a_s=(a_{s,B}\trans,a_{s,A},a_{s,h})\trans$. Restricting the residual representation to $D'$ gives
\[
v_{s,D'}
=eA_{D'}-B_{D'}a_{s,B}-A_{D'}a_{s,A}
-ea_{s,h}\ind_{I_{D'}}.
\]
Here $\ind_{I_{D'}}$ is the all-ones vector on $I_{D'}$. The common basis contains an intercept, so this last vector is in $\col(B_{D'})$. Thus
\[
v_{s,D'}\in\col([B_{D'},A_{D'}]).
\]
The use of the same basis at every node ensures that the restricted columns $B_{D'}$ are exactly the columns used for a candidate $r$ in $D'$. Applying Step 1 to that candidate yields
\[
[B_{D'},A_{D'}]\trans W_{D'}v_{r,D'}=0.
\]
Hence $v_{s,D'}\trans W_{D'}v_{r,D'}=0$. To obtain the claimed full-sample identity, use the diagonal form of $W$ and the fact that $v_r$ is zero outside $I_{D'}$:
\[
v_s\trans Wv_r
=\sum_{i\in I_{D'}}W_{ii}v_{s,i}v_{r,i}
=v_{s,D'}\trans W_{D'}v_{r,D'}=0.
\]
This argument applies to every eligible candidate in $D'$, whether or not it is selected in a grown tree.

\emph{Step 3: disjoint regions and normalized directions.}
If two candidates have disjoint parent regions, their zero-extended residuals have disjoint supports. Diagonality of $W$ then gives $v_s\trans Wv_r=0$ directly from the same coordinate sum. For both nested and disjoint cases, the definitions and $WVW=W$ imply
\[
\ell_s\trans V\ell_r
=\frac{v_s\trans WVWv_r}{d_sd_r}
=\frac{v_s\trans Wv_r}{d_sd_r}
=u_s\trans u_r=0.
\]
For each individual candidate, the normalization $d_q^2=v_q\trans Wv_q$ also gives
\[
u_q\trans u_q=\ell_q\trans V\ell_q
=\frac{v_q\trans Wv_q}{d_q^2}=1.
\]

\emph{Step 4: orthonormality within a fixed tree.}
Now fix a feasible labeled grown tree $H$. Any two distinct internal nodes are either in an ancestor--descendant relationship, in which the descendant parent lies within one child of the ancestor split, or have disjoint parent regions. Steps 2 and 3 therefore show that $\{u_w:w\in I(H)\}$ is orthonormal. For a node $w$ with parent region $Q_w$, Step 1 and zero extension also give, with $C_0=[B,A]$,
\[
u_w\trans V^{-1/2}C_0
=\frac{v_w\trans WC_0}{d_w}
=\frac{v_{w,Q_w}\trans W_{Q_w}[B_{Q_w},A_{Q_w}]}{d_w}
=0.
\]
Thus every whitened interaction direction is orthogonal to the stated common main-effect space.

\emph{Step 5: covariance of the fixed-tree linear scores.}
Under Assumption~\ref{pr:ass:gaussian}, the linear scores $S_w=\ell_w\trans Y$ for these fixed labels are jointly Gaussian, with means $\ell_w\trans\mu=d_w\theta_w$ and
\[
\operatorname{Cov}(S_w,S_{w'}\mid\mathcal D)
=\sigma^2\ell_w\trans V\ell_{w'}
=\begin{cases}
\sigma^2,&w=w',\\
0,&w\ne w'.
\end{cases}
\]
In particular, each linear score has variance $\sigma^2$, and distinct scores are independent under this unselected Gaussian law. This variance statement concerns $S_w$, rather than the squared split score $G_w(Y)=S_w^2$. Fixing $H$ here specifies the contrast vectors; it does not condition on the event $\widehat H(Y)=H$. No independence after tree selection is asserted.

Steps 1--4 are deterministic and require neither Gaussian sampling nor a causal mean representation. They use the positive diagonal covariance shape and the common basis containing an intercept. Gaussian sampling is used only in Step 5 for the distributional conclusion.
\end{proof}

\subsection{Proof of Lemma~\ref{pr:lem:completion}}
\begin{proof}
Fix the design and a feasible labeled path $\gamma$ ending at the target $\star$. Its contrast vectors, candidate rules, and node regions are then fixed. For Steps 1--4, fix $z\in\eta^\perp$ and $t\in\mathcal T_\gamma^{\rm grow}(z)$, and abbreviate $y=y_z(t)=z+vt$, since $c=v$ in~\eqref{pr:eq:decomposition}. By the definition of the growth section, ordinary growth at $y$ selects every prescribed split along $\gamma$.

\emph{Step 1: invariance of every off-path candidate score.}
An off-path region reached in the canonical construction lies either in a child of the target or in a sibling branch leaving its ancestral path. In the latter case it is disjoint from the target's parent. For every eligible candidate $r$ in any such region, Lemma~\ref{pr:lem:orthogonality} therefore gives
\[
\ell_r\trans v=\frac{v_r\trans Wv}{d_r}=0.
\]
Consequently, both its linear score and its squared split score agree at $y$ and $z$:
\[
\ell_r\trans y=\ell_r\trans z+t\ell_r\trans v=\ell_r\trans z,
\qquad
G_r(y)=G_r(z).
\]
This assertion concerns all candidates in the region, including candidates that are not ultimately selected. Thus every pairwise score comparison, with its exact equality status, is unchanged there.

\emph{Step 2: equality of the complete labeled grown trees.}
Compare ordinary growth at $y$ with the construction of $H_\gamma(z)$, proceeding from the root down. Suppose the two constructions have reached the same region. Eligibility, the candidate set, and the stopping rule coincide, because Assumption~\ref{pr:ass:growth} makes them functions only of the design and that region. If the node is on $\gamma$, ordinary growth chooses the prescribed split because $t$ belongs to the growth section; the canonical construction forces that same split. If the node is off the path, Step 1 shows that every candidate has the same score in the two constructions. Their fixed priorities therefore select the same winner, including at ties. In either case the chosen split labels and the resulting child regions coincide, or both constructions stop. Induction up to the finite maximum depth proves
\[
\widehat H(y_z(t))=H_\gamma(z).
\]
The restriction to the growth section is essential: competing candidates at a path node can have scores that vary with $t$, so the prescribed path need not be selected outside that section.

\emph{Step 3: invariance of all selected non-target scores.}
Consider an internal node $w\ne\star$ of this common tree. If it is off the path, Step 1 already gives $G_w(y)=G_w(z)$. Otherwise, it is a prescribed ancestor of the target. Applying the nested-region identity of Lemma~\ref{pr:lem:orthogonality} to that ancestor and the target yields
\[
\ell_w\trans v=\frac{v_w\trans Wv}{d_w}=0.
\]
The same linear-score calculation as in Step 1 therefore gives $G_w(y)=G_w(z)$ for every selected non-target node. This does not assert invariance of competing candidates at ancestor nodes.

\emph{Step 4: the target score.}
For the target, $\ell=d\eta$ and $\eta\trans z=0$. The normalization in~\eqref{pr:eq:contrast} gives
\[
\eta\trans v=\frac{v\trans Wv}{d^2}=1.
\]
It follows that
\[
\ell\trans y_z(t)=d\eta\trans(z+vt)=dt,
\qquad
G_\star(y_z(t))=d^2t^2.
\]
Together with Steps 2 and 3, this proves all three identities in~\eqref{pr:eq:completion}.

\emph{Step 5: measurability of the completion maps.}
The finite maximum depth and the finite candidate sets at every possible region give finitely many feasible labeled grown trees. Feasibility of $\gamma$ ensures that its forced nodes respect the design-only eligibility and stopping rules independently of $z$. For each fixed region and candidate, the loading $\ell_r$ is fixed by the design, and $G_r(z)=(\ell_r\trans z)^2$ is a quadratic polynomial in $z$. Hence each exact winning condition is a finite intersection of strict or weak polynomial inequalities, and is Borel measurable. For a fixed feasible tree $H$, the event $\{z:H_\gamma(z)=H\}$ is obtained by combining these off-path winning conditions with the forced path and the design-only stopping rules. It is therefore Borel. This proves measurability of $H_\gamma$ on $\eta^\perp$.

On each such fixed $H$, the pruning recursion in~\eqref{pr:eq:dp} forms continuation values by finitely many additions and maxima of continuous scores. Its decisions to retain or collapse an off-path node, including collapse at equality, are consequently Borel measurable. Joining these pruned hanging subtrees to the forced path proves measurability of $P_\gamma$ as well. This argument applies to every $z\in\eta^\perp$, even when its growth section is empty. It does not assert that the forced path survives ordinary pruning. All arguments above are deterministic; Gaussian sampling and a causal mean representation are not required.
\end{proof}

\subsection{Proof of Theorem~\ref{pr:thm:gate}}
\begin{proof}
Fix the design and a feasible labeled path $\gamma=(w_0,\ldots,w_k)$, with root $w_0$ and target $w_k=\star$. Under Assumption~\ref{pr:ass:pruning}, pruning can only collapse nodes and their descendants. Thus retention of $\gamma$ requires that $\gamma$ first occur in the grown tree. We determine the additional condition for retention on each growth section. The penalty $\lambda$ remains fixed throughout these calculations.

\emph{Step 1: fix the grown tree and the hanging-subtree contributions.}
Take $z\in\eta^\perp$ and $t\in\mathcal T_\gamma^{\rm grow}(z)$, and write $y=y_z(t)$, $x=d^2t^2$, and $H=H_\gamma(z)$. Lemma~\ref{pr:lem:completion} gives $\widehat H(y)=H$, target score $G_\star(y)=x$, and $G_w(y)=G_w(z)$ at every other internal node of $H$.

Every internal node in a hanging off-path subtree is a non-target node, so all scores in that subtree are the same at $y$ and $z$. At its terminal nodes the pruning values are zero in both cases. Moving upward, the recursion~\eqref{pr:eq:dp} applies the same additions and maxima to the same child values and node scores. Induction through the finite subtree therefore gives
\[
F_q^H(y)=F_q^H(z)=F_q^\gamma(z)
\]
at every node $q$ of each hanging subtree. In particular, the contributions from the target's two children and from every ancestor's sibling branch are functions of $z$ alone. The selected ancestor scores $G_{w_j}(z)$ are also fixed on the section.

\emph{Step 2: write the exact recursion along the path.}
Abbreviate the continuation and optimal pruning values on $H$ by
\[
V_j(t)=V_{w_j}^H(y),\qquad
F_j(t)=F_{w_j}^H(y)=\max\{0,V_j(t)\}.
\]
Step 1 and the definition of $a_k$ give
\[
\begin{aligned}
V_k(t)&=x-\lambda+F_{w_{k,L}}^\gamma(z)+F_{w_{k,R}}^\gamma(z)
=x+a_k(z),\\
V_j(t)&=G_{w_j}(z)-\lambda+F_{j+1}(t)+F_{b_j}^\gamma(z),
\qquad j=k-1,\ldots,0.
\end{aligned}
\]
We keep the maximum defining $F_{j+1}(t)$ until the sign of the corresponding continuation value has been established. In the subsequent top-down reconstruction, the target is reached and retained exactly when
\[
V_j(t)>0\qquad\text{for every }j=0,\ldots,k.
\]
Indeed, any nonpositive value collapses that path node and removes all descendants, whereas positive values at every path node permit reconstruction from the root through the target.

\emph{Step 3: necessity of the path inequalities.}
Suppose pruning retains the entire path. Step 2 gives $V_j(t)>0$ and hence $F_j(t)=V_j(t)$ for every $j$. At the target, $V_k(t)=x+a_k(z)>0$. If, for some $j<k$, the already evaluated child satisfies $F_{j+1}(t)=x+a_{j+1}(z)$, then
\[
\begin{aligned}
V_j(t)
&=G_{w_j}(z)-\lambda+x+a_{j+1}(z)+F_{b_j}^\gamma(z)\\
&=x+a_j(z)>0,
\end{aligned}
\]
where the last equality is~\eqref{pr:eq:a-ancestor}. Since this value is positive, $F_j(t)=x+a_j(z)$ as well. Backward induction from $k$ to $0$ proves the necessary inequalities $x+a_j(z)>0$ at every path node.

\emph{Step 4: sufficiency of the path inequalities.}
Conversely, suppose $x+a_j(z)>0$ for all $j=0,\ldots,k$. The target recursion gives $V_k(t)=F_k(t)=x+a_k(z)>0$. Applying the same backward calculation as in Step 3 shows successively that
\[
V_j(t)=F_j(t)=x+a_j(z)>0\qquad(j=k-1,\ldots,0).
\]
The top-down reconstruction therefore retains the root and then each successive node on $\gamma$, including the target. Choices within hanging branches cannot remove these retained path nodes. This proves both directions of the retention criterion; when $k=0$, only the target recursion is needed.

\emph{Step 5: combine the inequalities and account for equality.}
The finitely many strict inequalities are equivalent to
\[
x>\max_{0\le j\le k}\{-a_j(z)\}=\delta_\gamma(z).
\]
Their strictness follows from the stipulated collapse rule at continuation value zero. More explicitly, if $x=\delta_\gamma(z)\ge0$, then $x+a_j(z)\ge0$ for every $j$, with equality for at least one index. The backward recursion remains valid at these nonnegative values because $\max\{0,x+a_j(z)\}=x+a_j(z)$. Thus at least one path continuation value is zero, and the first such node reached from the root collapses. The target is consequently absent. This also shows why the boundary cannot be included in the pruning-survival gate.

\emph{Step 6: obtain the event identity and its sections.}
For an arbitrary outcome vector $y$, set $T(y)=\eta\trans y$ and $Z(y)=y-vT(y)$. The normalization $\eta\trans v=1$ gives $\eta\trans Z(y)=0$ and $y=y_{Z(y)}(T(y))$. If $y\notin E_\gamma^{\rm grow}$, pruning cannot create the missing path. Otherwise Steps 1--5 apply with $z=Z(y)$ and $t=T(y)$. They give the pointwise identity~\eqref{pr:eq:ret-event}. Along the line $y_z(t)$, these same identities give $Z(y_z(t))=z$ and $T(y_z(t))=t$. Taking the section at each fixed $z$ proves~\eqref{pr:eq:ret-section}, including the case of an empty growth section.

Finally, $d>0$ and $d^2t^2\ge0$. If $\delta_\gamma(z)<0$, the pruning inequality holds for every $t$; if it is zero, it holds exactly when $t\ne0$; and if it is positive, it holds exactly when $|t|>\sqrt{\delta_\gamma(z)}/d$. These are the three stated gate forms, each to be intersected with the growth section. The argument is deterministic and uses neither Gaussian sampling nor a causal mean representation.
\end{proof}

\subsection{Proof of Proposition~\ref{pr:prop:geometry}}
\begin{proof}
Fix the design, a feasible labeled path $\gamma$, and $z\in\eta^\perp$. The contrasts and candidate comparisons associated with that path are fixed. We characterize all $t\in\R$ along $y_z(t)=z+ct$; membership in the growth section will be determined by the comparisons themselves.

\emph{Step 1: express each winning comparison as a polynomial inequality.}
For any candidate $q$ at a path node, put $a_q=\ell_q\trans z$ and $b_q=\ell_q\trans c$. Its score along the line is
\[
G_q(y_z(t))=(a_q+b_qt)^2.
\]
For the prescribed winner $s$ and a competitor $r$, subtracting the two scores gives the polynomial
\[
\begin{aligned}
p_{s,r}(t)
&=(a_s+b_st)^2-(a_r+b_rt)^2\\
&=\{a_s-a_r+(b_s-b_r)t\}
  \{a_s+a_r+(b_s+b_r)t\}.
\end{aligned}
\]
This is~\eqref{pr:eq:quadratic} and has degree at most two. By~\eqref{pr:eq:win}, its required sign is $p_{s,r}(t)>0$ if $r$ has higher priority than $s$, and $p_{s,r}(t)\ge0$ if $s$ has higher priority. Intersecting these exact comparisons at all path nodes gives $\mathcal T_\gamma^{\rm grow}(z)$.

\emph{Step 2: handle constant comparisons.}
Some comparison polynomials may be constant. A positive constant satisfies either required inequality everywhere, and a negative constant satisfies neither anywhere. For a zero constant, the weak inequality is always true and the strict inequality is always false. Thus a constant comparison is either redundant or makes the section empty. It contributes no finite boundary point. This treatment preserves the specified tie rule even when two candidates have identical scores along the entire line.

\emph{Step 3: enumerate the finite boundary candidates.}
Theorem~\ref{pr:thm:gate} adds the strict pruning inequality
\[
h(t)=d^2t^2-\delta_\gamma(z)>0.
\]
Let $\mathcal B$ be the union of the real roots of every nonconstant comparison polynomial and of $h$. There are $m_\gamma$ comparisons, each with at most two real roots, and $h$ has at most two because $d>0$. Consequently,
\[
|\mathcal B|\le 2m_\gamma+2.
\]
We count distinct roots only. Linear comparisons, repeated roots, and the single root of $h$ when $\delta_\gamma(z)=0$ therefore cause no difficulty. If $\delta_\gamma(z)<0$, $h$ has no real root.

\emph{Step 4: recover the exact section from intervals and boundary points.}
Sort $\mathcal B$ increasingly. Its complement consists of finitely many open intervals, including the two unbounded tails; if $\mathcal B$ is empty, the only interval is $\R$. On each interval every nonconstant polynomial is continuous and nonzero, so its sign is constant. The constant comparisons also have fixed truth values. Testing all required inequalities at one point in each interval therefore determines membership throughout that interval.

Test each point of $\mathcal B$ separately, using the exact strict or weak winning comparisons and the strict pruning inequality. Taking all passing intervals and all passing boundary points gives precisely $\mathcal T_{\gamma,\lambda}^{\rm ret}(z)$. This procedure handles open or closed endpoints and isolated accepted points, including those arising at repeated roots. Merging adjacent accepted pieces gives a finite union of intervals, possibly with isolated points. Every finite boundary lies in $\mathcal B$, proving the claimed boundary bound.

\emph{Step 5: retain the complete set for inference.}
The construction tests the entire real line, so it recovers every component, including components not containing the observed statistic. Dropping such a component would change the selection section used in the conditional distribution. All accepted pieces must therefore be retained. The sharper two-interval description for the matching-GLS construction is established separately in Corollary~\ref{gls:cor:two-components}.

\emph{Step 6: reuse the observed grown tree and pruning calculations.}
Suppose the target survives at an observed outcome $y$, and set $t_{\rm obs}=\eta\trans y$ and $z_{\rm obs}=y-ct_{\rm obs}$. Survival implies that the path was grown, so $t_{\rm obs}\in\mathcal T_\gamma^{\rm grow}(z_{\rm obs})$. Lemma~\ref{pr:lem:completion} then gives
\[
H_\gamma(z_{\rm obs})=\widehat H(y),\qquad
G_w(z_{\rm obs})=G_w(y)\quad(w\ne\star)
\]
for the selected internal nodes of this tree. Thus the grown tree and its cached non-target scores are available without unconstrained regrowth at $z_{\rm obs}$.

Every hanging subtree contains only non-target nodes. A bottom-up application of~\eqref{pr:eq:dp} to their cached scores computes its pruning values at $z_{\rm obs}$. These subtrees are disjoint, so one traversal suffices. The cached ancestor scores and the resulting child and sibling values then evaluate~\eqref{pr:eq:a-target}--\eqref{pr:eq:delta} in one backward pass along the path. This justifies reuse of the scores and the stated finite computation of the pruning threshold once those scores are available.
\end{proof}

\subsection{Proof of Theorem~\ref{pr:thm:pivot}}
\begin{proof}
Fix the design $\mathcal D$ and the path index $\gamma$ in the theorem, and abbreviate $E=E_\gamma^{\rm ret}$. Thus $\Prob(E\mid\mathcal D)>0$, and $\eta$, $v$, and $d>0$ are fixed before conditioning on $E$. Write $J(z)=\mathcal T_{\gamma,\lambda}^{\rm ret}(z)$. We first obtain the selected conditional law and then establish each inferential guarantee.

\emph{Step 1: the unselected Gaussian decomposition.}
The definitions in~\eqref{pr:eq:contrast}, with zero extension outside the target parent, give
\[
V\eta=\frac{v}{d^2},\qquad
\eta\trans V\eta=\frac{v\trans Wv}{d^4}=\frac1{d^2}.
\]
Under Assumption~\ref{pr:ass:gaussian}, it follows that $T=\eta\trans Y$ is normal with mean $\theta=\eta\trans\mu$ and variance $\omega^2=\sigma^2/d^2>0$. Since $Z=Y-vT$, the pair $(T,Z)$ is jointly Gaussian, and
\[
\operatorname{Cov}(T,Z\mid\mathcal D)
=\sigma^2\eta\trans V-\omega^2v\trans
=\frac{\sigma^2}{d^2}v\trans-\frac{\sigma^2}{d^2}v\trans=0.
\]
Joint Gaussianity therefore implies independence of $T$ and $Z$ under the unselected law. This implication also holds for a singular Gaussian vector $Z$. All conditioning below uses regular conditional distributions; a full-dimensional Lebesgue density for $Z$ is unnecessary.

\emph{Step 2: positive section probability on the selected event.}
Theorem~\ref{pr:thm:gate} and the decomposition~\eqref{pr:eq:decomposition} identify the event pointwise as $E=\{T\in J(Z)\}$. Define
\[
q_\theta(z)=\int_{J(z)}\phi_\omega(t-\theta)\,dt.
\]
The selection event is measurable, as is $(z,t)\mapsto z+vt$, so this integral is measurable in $z$. Independence from Step 1 gives $q_\theta(z)=\Prob(E\mid Z=z,\mathcal D)$ for almost every $z$ under the unselected distribution. Consequently,
\[
\Prob\bigl(E\cap\{q_\theta(Z)=0\}\mid\mathcal D\bigr)
=\E\!\left[\ind\{q_\theta(Z)=0\}q_\theta(Z)\mid\mathcal D\right]=0.
\]
Thus $q_\theta(z)>0$ for almost every $z$ under the selected distribution of $Z$. For each such $z$, $J(z)$ has positive Lebesgue measure. Since $\phi_\omega(t-a)>0$ for every real $t,a$, the normalizing integral over this same $J(z)$ is positive for every candidate mean $a\in\R$.

\emph{Step 3: the selected conditional distribution.}
For any Borel set $B\subseteq\R$ and almost every selected $z$, conditioning the independent normal variable $T$ on $T\in J(z)$ gives
\[
\Prob(T\in B\mid Z=z,E,\mathcal D)
=\frac{\displaystyle\int_{B\cap J(z)}\phi_\omega(t-\theta)\,dt}
       {q_\theta(z)}.
\]
This is precisely the truncated-normal law in~\eqref{pr:eq:law}, with CDF $F^{J(z)}_{\theta,\omega}$. The normalizing integral includes every component of $J(z)$. Here $\theta$ is the mean parameter of the untruncated normal law, rather than the mean after selection.

\emph{Step 4: uniformity of the selective pivot.}
The law in Step 3 has a Lebesgue density and hence a continuous CDF. Its support may be disconnected; the flat portions of the CDF across gaps carry no probability and do not prevent the probability-integral transform. Therefore, with $U_\theta=F^{J(Z)}_{\theta,\omega}(T)$,
\[
\Prob(U_\theta\le u\mid Z=z,E,\mathcal D)=u,
\qquad 0\le u\le1,
\]
for almost every selected $z$. Averaging over that selected distribution gives
\[
\Prob(U_\theta\le u\mid E,\mathcal D)
=\E\!\left[\Prob(U_\theta\le u\mid Z,E,\mathcal D)
            \mid E,\mathcal D\right]=u.
\]
This proves both the conditional-on-$Z$ uniformity and~\eqref{pr:eq:pivot}.

\emph{Step 5: exact conditional size of the two-sided test.}
Under the fixed null hypothesis $\theta=\theta_0$, put $U_0=F^{J(Z)}_{\theta_0,\omega}(T)$. Step 4 applies to $U_0$ because its mean argument is then the true parameter. For $0\le r\le1$, the event $2\min\{U_0,1-U_0\}\le r$ is the union of $\{U_0\le r/2\}$ and $\{U_0\ge1-r/2\}$. These are disjoint up to endpoints of probability zero, so
\[
\Prob(p_{\theta_0}\le r\mid Z=z,E,\mathcal D)
=\frac r2+\frac r2=r.
\]
Taking $r=\alpha$ proves exact conditional rejection probability $\alpha$ for almost every selected $z$. Averaging over selected $Z$ proves the same equality conditional on $(E,\mathcal D)$.

\emph{Step 6: coverage by inversion.}
For a realized outcome, keep its contrast and section $J(Z)$ fixed while varying the candidate mean $a$ in~\eqref{pr:eq:ci}. The positive normalizing integrals from Step 2 make these CDFs well defined for every $a$. At the true parameter,
\[
\{\theta\in\mathcal C_{\gamma,\lambda}(Y)\}
=\{\alpha/2\le U_\theta\le1-\alpha/2\}.
\]
Uniformity from Step 4 gives probability $1-\alpha$ for this event conditional on $(Z=z,E,\mathcal D)$ for almost every selected $z$. Averaging once more yields coverage $1-\alpha$ conditional on $(E,\mathcal D)$. This completes the projection-target guarantees; no causal mean representation is used in the proof.
\end{proof}

\subsection{Proof of Corollary~\ref{pr:cor:endpoints}}
\begin{proof}
Fix the design and path index in Theorem~\ref{pr:thm:pivot}, and write $E=E_\gamma^{\rm ret}$. We work on a probability-one set of selected observations and then invert with their section and observed statistic held fixed.

\emph{Step 1: an interior observation on almost every selected section.}
Theorem~\ref{pr:thm:pivot} gives, for almost every $z$ under the distribution of $Z\mid(E,\mathcal D)$, a section $J=\mathcal T_{\gamma,\lambda}^{\rm ret}(z)$ with positive normal probability and a conditional law of $T$ having a Lebesgue density on $J$. Fix such a $z$. Proposition~\ref{pr:prop:geometry} shows that $J$ is a finite union of intervals, possibly with finitely many isolated points. Its boundary is therefore finite, so the conditional probability that $T$ lies on that boundary is zero. Since $T\in J$ on $E$, almost every conditional observation $t$ lies in the interior of $J$. In particular, $J$ has positive Lebesgue measure on both sides of $t$. For Steps 2--6, fix any such $J$ and $t$, together with the known $\omega>0$. The candidate mean $a$ ranges over all of $\R$; it is not restricted to $J$.

\emph{Step 2: a normalized exponential-tilt representation.}
Completing the square gives
\[
\phi_\omega(w-a)
=\exp\left\{\frac{aw}{\omega^2}-\frac{a^2}{2\omega^2}\right\}\phi_\omega(w).
\]
The factor depending only on $a$ cancels from the truncated density. Thus, writing
\[
M(a)=\int_J e^{aw/\omega^2}\phi_\omega(w)\,dw,
\qquad
f_a(w)=\frac{\ind\{w\in J\}e^{aw/\omega^2}\phi_\omega(w)}{M(a)},
\]
we have $0<M(a)<\infty$ for every finite $a$. Positivity follows from the positive measure of $J$, and finiteness follows from the Gaussian exponential moment. Define
\[
H(a)=F^J_{a,\omega}(t)=\int_{-\infty}^t f_a(w)\,dw.
\]
On any bounded range $|a|\le A$, the unnormalized integrands and their first derivatives in $a$ are bounded by a constant multiple of
\[
(1+|w|)e^{A|w|/\omega^2}\phi_\omega(w),
\]
which is integrable. Differentiation under the integral and continuity are therefore justified. Since $M(a)>0$, the ratio defining $H$ is continuously differentiable.

\emph{Step 3: strict decrease in the candidate mean.}
Let $Q_a$ have density $f_a$ and denote expectation under this law by $\E_a$. Differentiating the normalized density gives
\[
H'(a)=\frac{1}{\omega^2}
\operatorname{Cov}_a\bigl(\ind\{Q_a\le t\},Q_a\bigr).
\]
Put $p_a=H(a)$ and
\[
m_{-,a}=\E_a[Q_a\mid Q_a\le t],
\qquad
m_{+,a}=\E_a[Q_a\mid Q_a>t].
\]
The positive measure on both sides of $t$ and the strictly positive density on $J$ imply $0<p_a<1$ and $m_{-,a}<t<m_{+,a}$. These means are finite by the Gaussian tails. Consequently,
\[
H'(a)=\frac{p_a(1-p_a)}{\omega^2}
\bigl(m_{-,a}-m_{+,a}\bigr)<0.
\]
This argument applies even when $J$ is disconnected.

\emph{Step 4: the limits at both ends of the mean parameter space.}
Because $t$ is interior to $J$, choose $\varepsilon>0$ and bounded intervals of positive length
\[
K_-\subset J\cap(-\infty,t-\varepsilon],
\qquad
K_+\subset J\cap[t+\varepsilon,\infty).
\]
Set $b_\pm=\int_{K_\pm}\phi_\omega(w)\,dw>0$. For $a\ge0$, bound the numerator of $H(a)$ using $w\le t$ and the denominator using $K_+$:
\[
0\le H(a)
\le\frac{e^{at/\omega^2}}{b_+e^{a(t+\varepsilon)/\omega^2}}
=b_+^{-1}e^{-a\varepsilon/\omega^2}\longrightarrow0
\quad(a\to+\infty).
\]
For $a\le0$, apply the same comparison to the numerator of $1-H(a)$, where $w>t$, and to the denominator over $K_-$:
\[
0\le1-H(a)
\le\frac{e^{at/\omega^2}}{b_-e^{a(t-\varepsilon)/\omega^2}}
=b_-^{-1}e^{a\varepsilon/\omega^2}\longrightarrow0
\quad(a\to-\infty).
\]
Hence $H(a)\to1$ as $a\to-\infty$ and $H(a)\to0$ as $a\to+\infty$. The bounds apply to bounded or disconnected sections as well.

\emph{Step 5: existence, uniqueness, and ordering of the endpoints.}
Steps 2--4 show that $H$ is continuous and strictly decreasing, with limiting values one and zero. It therefore takes each value in $(0,1)$ exactly once. Since $0<\alpha<1$, the two equations
\[
H(L)=1-\alpha/2,
\qquad H(U)=\alpha/2
\]
have unique finite solutions. Strict decrease and $1-\alpha/2>\alpha/2$ give $L<U$. These are precisely the equations in~\eqref{pr:eq:endpoints}.

\emph{Step 6: identify the confidence set and return to the selected law.}
Strict decrease also gives
\[
\frac\alpha2\le H(a)\le1-\frac\alpha2
\quad\Longleftrightarrow\quad
L\le a\le U.
\]
Thus~\eqref{pr:eq:ci} is the closed interval $[L,U]$. Step 1 makes this argument valid with conditional probability one for almost every selected $z$. Averaging over $Z\mid(E,\mathcal D)$ proves the assertion almost surely on $E$. Coverage remains the guarantee of Theorem~\ref{pr:thm:pivot}. The argument establishes finite endpoints for these observations, but provides neither a common bound on their widths nor a finite expected width.
\end{proof}

\subsection{Proof of Corollary~\ref{pr:cor:address}}
\begin{proof}
Work under the assumptions of Theorem~\ref{pr:thm:pivot}. Fix the design $\mathcal D$ and the node address specified before observing $Y$, and let $\mathcal V$ be its availability event in the corollary. The split labels at that address and along its ancestral path remain outcome-dependent.

\emph{Step 1: enumerate the possible labeled paths.}
Let $\Gamma_{\rm addr}$ be the set of all design-feasible labeled paths ending at the specified address, including the target split label. The root has finitely many candidates, and each choice fixes the child region in which the next candidate set is finite. Repeating this argument up to the finite maximum depth shows that $\Gamma_{\rm addr}$ is finite. For each $\gamma\in\Gamma_{\rm addr}$, abbreviate $E_\gamma=E_\gamma^{\rm ret}$ and $C_\gamma(Y)=\mathcal C_{\gamma,\lambda}(Y)$.

\emph{Step 2: partition the availability event.}
If the specified address is internal in the final tree, its ancestors and its own split determine one labeled path $\gamma\in\Gamma_{\rm addr}$. Conversely, retention of such a path makes that address internal. The deterministic growth priorities and pruning convention give a unique labeled tree for every outcome, including at ties. Distinct paths ending at the same address therefore cannot occur together, and
\[
\mathcal V=\bigcup_{\gamma\in\Gamma_{\rm addr}}E_\gamma,
\qquad
E_\gamma\cap E_{\gamma'}=\varnothing
\quad(\gamma\ne\gamma').
\]
This partition is a pointwise identity; it includes feasible paths whose retention events have probability zero.

\emph{Step 3: identify the target and coverage on each part.}
On $\mathcal V$, denote the unique selected path by $\widehat\gamma(Y)$ and write the selected target and reported interval as
\[
\Theta_{\rm sel}(Y)=\theta_{\widehat\gamma(Y)},
\qquad
C_{\rm sel}(Y)=C_{\widehat\gamma(Y)}(Y).
\]
Each $\theta_\gamma=\eta_\gamma\trans\mu$ is fixed for fixed design and path labels. The randomness in $\Theta_{\rm sel}$ comes from the selected index. On $E_\gamma$, the target is $\theta_\gamma$ and the interval is exactly $C_\gamma(Y)$.

Define
\[
\Gamma_+=\{\gamma\in\Gamma_{\rm addr}:
\Prob(E_\gamma\mid\mathcal D)>0\}.
\]
For every $\gamma\in\Gamma_+$, Theorem~\ref{pr:thm:pivot} gives
\[
\Prob\{\theta_\gamma\in C_\gamma(Y)\mid E_\gamma,\mathcal D\}
=1-\alpha.
\]
Here we use that theorem's guarantee conditional on the retained-path event after averaging over its own target-specific $Z$. No common conditioning vector across different paths is needed. The finitely many events with $\gamma\notin\Gamma_+$ have a probability-zero union, so their conditional probability given $\mathcal V$ is also zero because $\Prob(\mathcal V\mid\mathcal D)>0$.

\emph{Step 4: average over the disjoint parts.}
Using Steps 2 and 3 and the positive availability probability,
\[
\begin{aligned}
&\Prob\{\Theta_{\rm sel}(Y)\in C_{\rm sel}(Y)
\mid\mathcal V,\mathcal D\}\\
&\quad=
\frac{\displaystyle\sum_{\gamma\in\Gamma_+}
\Prob\{\theta_\gamma\in C_\gamma(Y),\ E_\gamma\mid\mathcal D\}}
{\Prob(\mathcal V\mid\mathcal D)}\\
&\quad=
\frac{(1-\alpha)\displaystyle\sum_{\gamma\in\Gamma_+}
\Prob(E_\gamma\mid\mathcal D)}
{\Prob(\mathcal V\mid\mathcal D)}
=1-\alpha.
\end{aligned}
\]
The last equality follows because the sum of the retention probabilities is $\Prob(\mathcal V\mid\mathcal D)$. These probabilities are used only to prove coverage and need not be computed by the procedure. This establishes coverage for the selected target at the one prespecified address whenever it is available.
\end{proof}

\subsection{Proof of Lemma~\ref{pr:lem:prunedcompletion}}
\begin{proof}
Fix the design, a feasible labeled path $\gamma=(w_0,\ldots,w_k)$ with target $w_k=\star$, and any $z\in\eta^\perp$ and $t\in\mathcal T_{\gamma,\lambda}^{\rm ret}(z)$. Write $y=y_z(t)=z+vt$, $x=d^2t^2$, and $H=H_\gamma(z)$. We show that actual pruning at $y$ agrees with the forced-path completion at $z$. All pruning values below are evaluated on the same fixed labeled tree $H$.

\emph{Step 1: fix the grown tree and its non-target scores.}
Theorem~\ref{pr:thm:gate} gives
\[
\mathcal T_{\gamma,\lambda}^{\rm ret}(z)
\subseteq\mathcal T_\gamma^{\rm grow}(z).
\]
Lemma~\ref{pr:lem:completion} therefore applies and yields
\[
\widehat H(y)=H,\qquad
G_\star(y)=x,\qquad
G_q(y)=G_q(z)\quad(q\in I(H),\ q\ne\star).
\]
This proves the grown-tree identity in~\eqref{pr:eq:prunedcompletion}. It also supplies the fixed non-target scores used in the remaining steps.

\emph{Step 2: equality of the recursion values in every hanging subtree.}
For each ancestor $w_j$, $j<k$, let $b_j$ be the sibling of $w_{j+1}$. The roots of the hanging off-path subtrees are
\[
\mathcal R_\gamma
=\{w_{k,L},w_{k,R}\}\cup\{b_j:0\le j<k\}.
\]
The sibling collection is empty when $k=0$. For $r\in\mathcal R_\gamma$, let $H_r$ be the subtree of $H$ rooted at $r$, allowing $H_r$ to be a single leaf. These subtrees are disjoint and exhaust the nodes of $H$ outside the path.

Every internal node $q$ of $H_r$ differs from the target, so Step 1 gives $G_q(y)=G_q(z)$. At a terminal node $q$, the initialization in~\eqref{pr:eq:dp} gives $F_q^H(y)=F_q^H(z)=0$. If the two child values agree at an internal node $q$, that recursion gives
\[
\begin{aligned}
V_q^H(y)
&=G_q(y)-\lambda+F_{q_L}^H(y)+F_{q_R}^H(y)\\
&=G_q(z)-\lambda+F_{q_L}^H(z)+F_{q_R}^H(z)
=V_q^H(z),\\
F_q^H(y)
&=\max\{0,V_q^H(y)\}
=\max\{0,V_q^H(z)\}=F_q^H(z).
\end{aligned}
\]
Induction from the leaves proves equality of $F$ at every node and of $V$ at every internal node of every $H_r$.

\emph{Step 3: equality of the locally pruned subtrees, including ties.}
Let $P_r^H(u)$ denote the rooted pruned tree obtained by starting the top-down reconstruction at $r$ on $H_r$, using the values evaluated at outcome $u$. This is a local pruning operation on the fixed $H_r$. For every $q\in I(H_r)$, Step 2 implies
\[
\ind\{V_q^H(y)>0\}=\ind\{V_q^H(z)>0\}.
\]
Starting at $r$, both reconstructions therefore make the same retain-or-collapse decision. When they retain a node, they reach its same two children and repeat the same decisions. When its continuation value is zero, both collapse it by the stipulated convention. Induction down the finite subtree proves
\[
P_r^H(y)=P_r^H(z)
\qquad(r\in\mathcal R_\gamma),
\]
as labeled rooted trees. This includes the case of a terminal hanging root.

\emph{Step 4: retention of the complete target path.}
Since $t$ belongs to the retained section, Theorem~\ref{pr:thm:gate} gives
\[
x>\delta_\gamma(z)=\max_{0\le j\le k}\{-a_j(z)\},
\qquad x+a_j(z)>0\quad(j=0,\ldots,k).
\]
Using the fixed hanging-subtree values from Step 2, backward induction in~\eqref{pr:eq:dp}, starting at the target, gives
\[
V_{w_j}^H(y)=F_{w_j}^H(y)=x+a_j(z)>0
\qquad(j=0,\ldots,k),
\]
as in the proof of Theorem~\ref{pr:thm:gate}. Thus actual reconstruction from the root retains every path node and reaches every hanging root. These path values may vary with $t$; only their positivity is used here. Constancy of the recursion values was needed only inside the hanging subtrees.

\emph{Step 5: assemble the same complete final tree.}
Let $I_\gamma=\{w_0,\ldots,w_k\}$. Step 4 and the decomposition into hanging subtrees give
\[
\begin{aligned}
I(\widehat P_\lambda(y))
&=I_\gamma\cup\bigcup_{r\in\mathcal R_\gamma}I(P_r^H(y))\\
&=I_\gamma\cup\bigcup_{r\in\mathcal R_\gamma}I(P_r^H(z))\\
&=I(P_\gamma(z)).
\end{aligned}
\]
The second equality is Step 3. The last is exactly Definition~\ref{pr:def:completion}: retain the path by force and attach the locally pruned hanging subtrees evaluated at $z$. Both final trees are obtained by collapsing nodes of the same labeled $H$. Their identical internal-node sets therefore determine identical terminal regions, split variables, thresholds, and child directions. Consequently,
\[
\widehat P_\lambda(y_z(t))=P_\gamma(z),
\]
which completes~\eqref{pr:eq:prunedcompletion}. The argument is deterministic and keeps the exact tie rules throughout, so the identities hold for every stated $z,t$, without a Gaussian or probability-one qualification. If the retained section is empty, its assertion is vacuous.
\end{proof}

\subsection{Proof of Theorem~\ref{pr:thm:fulltree}}
\begin{proof}
Fix the design, the feasible labeled path $\gamma$, and the penalty $\lambda$. Its contrast vectors $\eta,v$ are consequently fixed. Keep the labeled final tree $P$ in the statement fixed, and, for the pair assertion, also fix the specified grown tree $H$. Write $E=E_\gamma^{\rm ret}$ and $J(z)=\mathcal T_{\gamma,\lambda}^{\rm ret}(z)$. In Steps 1--4 we view selection events as subsets of the outcome space and prove pointwise identities. The Gaussian assumption is used only in Steps 5--6.

\emph{Step 1: canonical identities on the retained-path event.}
For any outcome vector $y$, let $T(y)=\eta\trans y$ and $Z(y)=y-vT(y)$. Since $\eta\trans v=1$,
\[
\eta\trans Z(y)=0,\qquad
y=y_{Z(y)}(T(y)).
\]
Membership in $E$ is therefore equivalent to $T(y)\in J(Z(y))$. Applying Lemma~\ref{pr:lem:prunedcompletion} on that section yields the pointwise implication
\[
y\in E\quad\Longrightarrow\quad
\begin{cases}
\widehat H(y)=H_\gamma(Z(y)),\\
\widehat P_\lambda(y)=P_\gamma(Z(y)).
\end{cases}
\]
The fixed comparison tree $H$ is distinct from the completion map $H_\gamma$, which can take different values as $z$ varies.

\emph{Step 2: both inclusions for the final-tree event.}
Suppose first that $y\in E_P$, so $\widehat P_\lambda(y)=P$. Because $P$ contains $\gamma$ as an internal path, $y\in E$. Step 1 then gives
\[
P_\gamma(Z(y))=\widehat P_\lambda(y)=P,
\qquad Z(y)\in\mathcal A_P^\gamma.
\]
Thus $E_P\subseteq E\cap\{Z\in\mathcal A_P^\gamma\}$. Conversely, if $y\in E$ and $Z(y)\in\mathcal A_P^\gamma$, Step 1 gives
\[
\widehat P_\lambda(y)=P_\gamma(Z(y))=P,
\]
so $y\in E_P$. These two inclusions prove~\eqref{pr:eq:factor-final} with the exact growth and pruning tie conventions.

\emph{Step 3: the grown/final-pair factorization and measurability.}
If $y\in E_{H,P}$, its final tree again contains $\gamma$, so $y\in E$. Both identities in Step 1 apply. Conversely, on $E$ those identities turn the two completion equalities into the actual grown- and final-tree equalities. Hence
\[
\begin{aligned}
E_{H,P}
&=\{y\in E:H_\gamma(Z(y))=H,\ P_\gamma(Z(y))=P\}\\
&=E\cap\{Z\in\mathcal A_{H,P}^\gamma\},
\end{aligned}
\]
which proves~\eqref{pr:eq:factor-pair} in both directions.

Lemma~\ref{pr:lem:completion} gives measurable completion maps with finitely many labeled-tree values. In particular,
\[
\begin{aligned}
\mathcal A_P^\gamma&=P_\gamma^{-1}(\{P\}),\\
\mathcal A_{H,P}^\gamma
&=H_\gamma^{-1}(\{H\})\cap P_\gamma^{-1}(\{P\})
\end{aligned}
\]
are Borel subsets of $\eta^\perp$. The retained-path event $E$ is measurable by its finite growth comparisons and pruning inequalities, and $Z$ is linear. Thus all factors and events in both factorizations are measurable.

\emph{Step 4: sections for every $z$ and compatibility of an observation.}
Write $K=P$ or $K=(H,P)$ and use its corresponding $E_K$ and $\mathcal A_K^\gamma$. Along $y_z(t)=z+vt$, with $z\in\eta^\perp$, we have
\[
T(y_z(t))=t,\qquad Z(y_z(t))=z.
\]
The factorizations in Steps 2--3 consequently give
\[
\begin{aligned}
\ind\{y_z(t)\in E_K\}
&=\ind\{z\in\mathcal A_K^\gamma\}\,
  \ind\{y_z(t)\in E\}\\
&=\ind\{z\in\mathcal A_K^\gamma\}\,
  \ind\{t\in J(z)\}.
\end{aligned}
\]
This proves~\eqref{pr:eq:section-case}: the full-event section equals $J(z)$ at a compatible $z$ and is empty otherwise. The identity holds for every $z$, including empty sections and exact ties. If an observed $y$ belongs to $E_K$, the factorization ensures $Z(y)\in\mathcal A_K^\gamma$. Its observed section therefore satisfies~\eqref{pr:eq:equal-sections}.

\emph{Step 5: transfer the conditional Gaussian law to the full event.}
Now impose Assumption~\ref{pr:ass:gaussian} and suppose $\Prob(E_K\mid\mathcal D)>0$. Since $E_K\subseteq E$, also $\Prob(E\mid\mathcal D)>0$, so Theorem~\ref{pr:thm:pivot} applies to the fixed path. Let $\nu_E$ and $\nu_K$ be the distributions of $Z$ conditional on $(E,\mathcal D)$ and $(E_K,\mathcal D)$, respectively. The factorization gives, for every Borel $B\subseteq\eta^\perp$,
\[
\nu_K(B)=
\frac{\nu_E(B\cap\mathcal A_K^\gamma)}
     {\nu_E(\mathcal A_K^\gamma)},
\qquad
\nu_E(\mathcal A_K^\gamma)
=\frac{\Prob(E_K\mid\mathcal D)}{\Prob(E\mid\mathcal D)}>0.
\]
Thus $\nu_K\ll\nu_E$ and $\nu_K(\mathcal A_K^\gamma)=1$. In particular, every exceptional set of $z$ excluded by Theorem~\ref{pr:thm:pivot} still has probability zero after selection of the full event.

For $\nu_K$-almost every $z$, the section $J(z)$ has positive normal probability and $z\in\mathcal A_K^\gamma$. Once this $z$ and $E$ are fixed, the additional event $\{Z\in\mathcal A_K^\gamma\}$ is automatic. With $\theta=\eta\trans\mu$ and $\omega^2=\sigma^2/d^2$, Theorem~\ref{pr:thm:pivot} therefore yields, for every Borel $C\subseteq\R$,
\[
\begin{aligned}
\Prob(T\in C\mid Z=z,E_K,\mathcal D)
&=\Prob(T\in C\mid Z=z,E,\mathcal D)\\
&=\frac{\displaystyle\int_{C\cap J(z)}\phi_\omega(s-\theta)\,ds}
        {\displaystyle\int_{J(z)}\phi_\omega(s-\theta)\,ds}.
\end{aligned}
\]
This is the same truncated-normal kernel as~\eqref{pr:eq:law}, now under the complete-final-tree or grown/final-pair event.

\emph{Step 6: inferential guarantees and numerical agreement.}
Put $U_\theta=F^{J(Z)}_{\theta,\omega}(T)$. The continuous-CDF argument in Theorem~\ref{pr:thm:pivot} applies to the kernel in Step 5, giving
\[
\Prob(U_\theta\le u\mid Z=z,E_K,\mathcal D)=u,
\qquad 0\le u\le1,
\]
for $\nu_K$-almost every $z$. Averaging over this selected distribution gives
\[
\Prob(U_\theta\le u\mid E_K,\mathcal D)
=\int u\,\nu_K(dz)=u.
\]
The same two-tail calculation and fixed-section inversion as in that theorem give
\[
\begin{aligned}
\Prob(p_{\theta_0}\le\alpha\mid E_K,\mathcal D)
&=\alpha\quad\text{when }\theta=\theta_0,\\
\Prob\{\theta\in\mathcal C_{\gamma,\lambda}(Y)
\mid E_K,\mathcal D\}
&=1-\alpha.
\end{aligned}
\]
These guarantees also hold conditional on $Z$ for $\nu_K$-almost every $z$.

Finally, at such a compatible observed section with $t_{\rm obs}=\eta\trans y$, Step 4 gives $\mathcal T_K(z)=J(z)$. Their positive measure makes the normalizing integrals positive for every candidate mean $a$, and
\[
F^{\mathcal T_K(z)}_{a,\omega}(t_{\rm obs})
=F^{J(z)}_{a,\omega}(t_{\rm obs})
\qquad(a\in\R).
\]
The contrast, reference variance, and observed statistic are the same in both calculations. Hence their numerical pivots, two-sided $p$-values, and inverted confidence intervals coincide. Steps 1--4 establish pointwise structural identities; the inferential conclusions in Steps 5--6 hold on the stated selected probability-one set.
\end{proof}

\subsection{Justification of Remark~\ref{pr:rem:simultaneous}}
\label{supp:proof:simultaneous}
\begin{proof}
For a fixed selected $P$, each $\eta_v$ is fixed by its labeled ancestral path and design. Theorem~\ref{pr:thm:fulltree} gives conditional super-uniformity (in fact uniformity) of its $p$-value for every true projection null. If $\mathcal I_0(P;\mu)$ denotes the set of those null indices, the union bound gives
\[
\Prob\left(\bigcup_{v\in\mathcal I_0(P;\mu)}\{p_v\le\alpha/K\}\,\middle|\,E_P,\mathcal D\right)
\le |\mathcal I_0(P;\mu)|\alpha/K\le\alpha.
\]
Applying the same bound to the $K$ individual noncoverage events proves simultaneous coverage of the adjusted intervals. For $K=0$ there is no rejection or noncoverage event to consider. No joint independence assertion is used.
\end{proof}

\section{Further structural results and limits}\label{supp:sec:extensions}
\subsection{Relation to existing selective tree inference}
\label{pr:sec:scope}

Interaction trees and covariate-adjusted interaction trees already use interaction-complexity criteria of the form $\sum G_w-\lambda|I(P)|$ \citep[Section 2.2]{Su2009}; see also \citet[Section 2]{Steingrimsson2019}. The present rule specifies known-variance GLS residual scores, a common outcome-independent basis, and fixed tuning. It does not claim a new pruning objective or coverage for every published CAIT implementation.

Tree-Values already gives branch growth and fixed-penalty survival constraints \citep[Section 4.2]{Neufeld2022}. Its Appendix D.3, Lemma 27 establishes grown-tree and off-path-pruning constancy under its perturbation condition; Appendix D.4, Lemma 30 reduces sibling-contrast pruning constraints to two tails. The Gaussian conditioning principle and quadratic sections also predate the present construction \citep{Lee2016}; see Loftus and Taylor (2015, \emph{Selective inference in regression models with groups of variables}, \href{https://arxiv.org/abs/1511.01478}{arXiv:1511.01478}).

Our structural extension uses weighted common-basis orthogonality for treatment-interaction residuals whose perturbations need not be childwise constant. Tree-Values also discusses the general known-covariance Gaussian decomposition in its Section 7, while noting that its computational results need not extend. Here positive \emph{diagonal} $V$ and matching GLS scores preserve the local restriction geometry. The explicit retained-path/final-tree factorization is established for this specified rule; arbitrary correlated covariance or a change only to standard errors is not covered by that proof.

\subsection{A weighted RSS representation for the selected directions}
Recall $C_0=[B,A]$ and the whitened vectors in Lemma~\ref{pr:lem:orthogonality}.

\begin{proposition}[An exact weighted residual-sum-of-squares interpretation]
\label{pr:prop:rss}
Fix a feasible grown tree $H$ and a subset $S\subseteq I(H)$. Let $D_S$ contain the original-coordinate mean directions $v_w/d_w$, and put $U_S=V^{-1/2}D_S$. Then
\begin{equation}\label{pr:eq:rss}
\begin{aligned}
P_{V^{-1/2}[C_0,D_S]}&=P_{V^{-1/2}C_0}+U_SU_S\trans,\\
\RSS_W(C_0;y)-\RSS_W([C_0,D_S];y)&=\sum_{w\in S}G_w(y),
\end{aligned}
\end{equation}
where $\RSS_W(C;y)=\norm{(I-P_{V^{-1/2}C})V^{-1/2}y}^2$.
\end{proposition}
\begin{proof}
Lemma~\ref{pr:lem:orthogonality} gives $U_S\trans U_S=I$ and $P_{V^{-1/2}C_0}U_S=0$. Orthogonal column spaces give the projector identity. Applying its quadratic form to $V^{-1/2}y$ gives the RSS identity because $u_w\trans V^{-1/2}y=\ell_w\trans y$.
\end{proof}
Thus the pruning scores are actual weighted RSS improvements in a regression augmented by these orthogonal interaction directions. The augmentation columns are $v_w/d_w$, not the score loadings $\ell_w=Wv_w/d_w$. This is not equality with an ordinary CART leaf-mean regression, a separately fitted prognosis in each leaf, or a population treatment-policy criterion.

\subsection{Score envelopes and ancestor costs}

\begin{proposition}[Score envelopes and the role of ancestors]
\label{pr:prop:envelopes}
For a subtree $P\preceq H_\gamma(z)$, put
\[
C_P(z)=\sum_{w\in I(P)\setminus\{\star\}}G_w(z)-\lambda|I(P)|.
\]
Then on a growth section,
\begin{equation}\label{pr:eq:envelope-decomp}
Q_\lambda(P;y_z(t))=C_P(z)+\ind\{\star\in I(P)\}d^2t^2.
\end{equation}
The best target-retaining and target-excluding constants satisfy
\begin{equation}\label{pr:eq:envelopes}
C_{\rm in}(z)=a_0(z),\qquad
C_{\rm out}(z)=\max_{0\le j\le k}\{a_0(z)-a_j(z)\},\qquad
C_{\rm out}(z)-C_{\rm in}(z)=\delta_\gamma(z).
\end{equation}
In particular, the survival condition is generally not $G_\star>\lambda$.
\end{proposition}
\begin{proof}
Equation~\eqref{pr:eq:envelope-decomp} follows from~\eqref{pr:eq:completion}. A target-retaining subtree keeps every path split, while each hanging subtree is optimized independently. Its best score is $d^2t^2+a_0(z)$. A target-excluding subtree has a first collapsed path node $w_j$, possibly the root or the target itself. Keeping the preceding path and optimizing its hanging branches gives score $a_0(z)-a_j(z)$. Maximizing over $j$ gives~\eqref{pr:eq:envelopes}.
\end{proof}

For example, suppose a grown tree has root gain $0.2$ and a single internal left child of gain $1.5$, with $\lambda=1$. Retaining the target child also requires retaining the root, yielding $0.2+1.5-2=-0.3$, below the score zero obtained by collapsing the root to a leaf. The child's gain exceeds $\lambda$, but it is pruned. Conversely, profitable descendants can support an ancestor whose own gain is below $\lambda$.

\subsection{Conditions for extension and their limits}
\label{pr:sec:limits}
The structural sufficient conditions for Theorem~\ref{pr:thm:fulltree} are finite local growth with design-only candidates and stopping rules, common-basis orthogonality, unchanged local contrasts for retained splits, and a pruning rule whose full output on a target-surviving line section is determined by $z$ alone. The specified additive fixed-penalty rule has this property by Lemma~\ref{pr:lem:prunedcompletion}.

\begin{proposition}[A broader fixed-family sufficient condition]
\label{pr:prop:fixed-family}
Replace Assumption~\ref{pr:ass:pruning} by the following rule. For each design-feasible grown $H$, let $\mathcal F_H$ be a nonempty finite family of rooted pruned subtrees determined by $H$ and the design alone. Let the finite real-valued penalty $\rho_H(P)$ and a total priority order on $\mathcal F_H$ be fixed given $H$ and the design, with no further outcome dependence. Select the priority-winning maximizer of
\[
Q(P;y)=\sum_{w\in I(P)}G_w(y)-\rho_H(P),\qquad P\in\mathcal F_H.
\]
For each canonical $H_\gamma(z)$, suppose both the target-retaining and target-excluding subfamilies are nonempty. Define
\[
C_P(z)=\sum_{w\in I(P)\setminus\{\star\}}G_w(z)-\rho_{H_\gamma(z)}(P),
\]
and let $P_{\rm in}(z),P_{\rm out}(z)$ be their priority-winning maximizers, with values $M_{\rm in}(z),M_{\rm out}(z)$. Put $\Delta(z)=M_{\rm out}(z)-M_{\rm in}(z)$. The exact retention gate is
\begin{equation}\label{pr:eq:general-gate}
\{t:d^2t^2>\Delta(z)\}
\ \cup\
\{t:d^2t^2=\Delta(z),\ P_{\rm in}(z)\prec P_{\rm out}(z)\}.
\end{equation}
Intersecting this gate with the growth section gives valid known-variance selective inference. On the retained section, the complete output is $P_{\rm in}(z)$, so the final-tree and grown/final-pair factorizations and compatible-section equivalence also hold, with this completion replacing $P_\gamma(z)$.
\end{proposition}
\begin{proof}
Lemma~\ref{pr:lem:completion} fixes the grown tree on a growth section. For every admissible $P$,
\[
Q(P;y_z(t))=C_P(z)+\ind\{\star\in I(P)\}d^2t^2.
\]
Thus rankings within each subfamily, including their priorities, depend only on $z$. Comparing their two best values gives~\eqref{pr:eq:general-gate}, with its exact tie rule. If the target survives, the overall winner is $P_{\rm in}(z)$. This supplies precisely the canonical-output identity used in the proof of Theorem~\ref{pr:thm:fulltree}; the same two inclusions prove both factorizations. The Gaussian conditioning proof in Theorem~\ref{pr:thm:pivot} applies to the new section. Empty subfamilies, if allowed, give an always-false or always-true retention gate and can be treated separately.
\end{proof}

This proposition accommodates a fixed leaf-count budget or other fixed restrictions for this additive objective, provided the stipulated subfamilies are nonempty; a target that cannot survive is unavailable. The simple recursion in~\eqref{pr:eq:dp} and threshold in~\eqref{pr:eq:delta} specifically require the unrestricted rooted-subtree family and collapse at equality. Constrained optimization may require a different dynamic program. An arbitrary fixed priority may include boundary points that the primary convention excludes.

\begin{remark}[Why a generic pruner need not give the equivalence]
\label{pr:rem:counterexample}
Consider a design with one prescribed candidate at every reached node, so the grown labeled tree is fixed. Let a different pruner always retain the root, but retain an off-path child split only when the root gain exceeds a fixed $\kappa>0$. For root-target inference, path growth and root survival impose no restriction. The complete final tree with that child retained additionally requires $d^2T^2>\kappa$. Its output therefore varies with $t$ on the root-surviving section, and the equivalence fails. Correct Gaussian inference is still possible if that full selection restriction is included. The additive pruning structure above, rather than measurability alone, supplies the stronger equivalence.
\end{remark}

\begin{remark}[Why growth-only inference fails after pruning]
\label{pr:rem:stump}
Consider a one-candidate stump under its projection null. Then $\ell\trans Y/\sigma\sim N(0,1)$, and with $\lambda=\sigma^2$ it survives only when $|\ell\trans Y|>\sigma$. For $0<\alpha<2\{1-\Phi(1)\}$, the ordinary two-sided level-$\alpha$ Gaussian rejection event is contained in this retention event. Its conditional rejection probability is therefore
\[
\frac{\alpha}{2\{1-\Phi(1)\}}>\alpha.
\]
Thus a correct growth-stage calculation that ignores survival need not remain valid after pruning.
\end{remark}

\subsection{Why changing only the OLS standard error is insufficient}
\label{gls:sec:ols-failure}
For an arbitrary fixed contrast $\eta$ and known Gaussian covariance $\Sigma$, the decomposition
\[
T=\eta\trans Y,\qquad
c=\frac{\Sigma\eta}{\eta\trans\Sigma\eta},\qquad Z=Y-cT
\]
always gives $T\perp Z$. Exact selective Gaussian inference is therefore possible using the \emph{actual} section of the whole selected event. This general fact does not ensure that the complete grown tree or off-path pruning is fixed along that line. For unweighted interaction directions $u_s$, Euclidean orthogonality need not imply $u_s\trans\Sigma u_r=0$.

\begin{proposition}[OLS structural reductions can fail under arm heteroskedasticity]
\label{gls:prop:ols-failure}
With a fixed common prognostic basis containing a nonconstant covariate, there are eligible ancestor and descendant interaction splits with unweighted residuals $v_s^{\rm O}$ and $v_r^{\rm O}$ satisfying
\[
(v_s^{\rm O})\trans v_r^{\rm O}=0,
\qquad
(v_s^{\rm O})\trans\Sigma v_r^{\rm O}\ne0,
\]
even for $\Sigma=\operatorname{diag}(1+A_i)$. For a local growth rule with these two prescribed splits, the target-retention section need not be a squared-score threshold, and the final-tree section need not equal the retained-path section. Thus changing only the OLS standard error does not give the GLS structural results.
\end{proposition}
\begin{proof}
Take $n=12$, $A=(0,1,0,1,0,1,0,1,0,1,0,1)\trans$, $B=[\ind,x]$, and
\[
x=(0,1,2,4,1,3,5,8,0,2,6,7)\trans.
\]
Let the root's left child contain observations $1,\ldots,8$, and split that child into observations $1,\ldots,4$ and $5,\ldots,8$. Additional baseline split coordinates can realize these geometric regions; they need not enter $B$. Use ordinary residuals of $Ah$ on $[B,A,h]$ in each parent, extending the descendant residual by zero. Both residuals are nonzero. Direct solution of the normal equations gives
\[
(v_s^{\rm O})\trans v_r^{\rm O}=0,
\qquad
(v_s^{\rm O})\trans\operatorname{diag}(1+A)v_r^{\rm O}
=\frac{16}{2915}\ne0.
\]
Give each reached splitting node only its prescribed candidate and make all other nodes terminal by design, so the grown labeled tree is fixed. For root-target inference set $d=\norm{v_s^{\rm O}}$, $u_s=v_s^{\rm O}/d$, $u_r=v_r^{\rm O}/\norm{v_r^{\rm O}}$, and use the correct covariance direction $c=\Sigma\eta/(\eta\trans\Sigma\eta)$ with $\eta=u_s/d$. Put $b=u_r\trans c\ne0$ and $z=\sqrt2u_r\in\eta^\perp$. Along $z+ct$,
\[
G_{\rm root}=d^2t^2,
\qquad G_{\rm child}=(\sqrt2+bt)^2.
\]
At $\lambda=1$, root retention is exactly
\[
d^2t^2+\max\{0,(\sqrt2+bt)^2-1\}>1.
\]
For sufficiently small positive and negative $t$, its truth values differ because its continuation value near zero is $(d^2+b^2)t^2+2\sqrt2bt$. No condition $d^2t^2>\delta(z)$ can have this asymmetry.

The final child also changes on root-retained sections. At $t=-\sqrt2/b$ its gain is zero while the root gain is $2d^2/b^2>1$. To verify the inequality, write $b=d\,u_r\trans\Sigma u_s/(u_s\trans\Sigma u_s)$, use the strict covariance Cauchy--Schwarz inequality, and note that the eigenvalues of $\Sigma$ are 1 and 2; hence $b^2<2d^2$. At sufficiently large $|t|$, both splits survive. Thus one compatible retained root section contains outcomes with different final trees, refuting automatic section equivalence for the unmodified OLS rule.
\end{proof}
The intercept-only nuisance space is a special exception: multiplication by an arm-specific variance preserves $\operatorname{span}(\ind,A)$. With additional common covariates, that closure need not hold. Matching GLS residualization avoids needing such an extra closure assumption.

\subsection{Known shape and unknown scale}
\label{gls:sec:known-shape}
The deterministic geometry and pruning identities use the known positive diagonal $V$, not the numerical value of $\sigma^2$. If only the variance ratio is known, the structural statements still hold with $\operatorname{Var}(T)=\sigma^2/d^2$, but the known-normal pivot is not available without the scale. Estimating either the scale or the shape and substituting it does not inherit the exact theorem. Conditioning on an independently estimated variance does not make its realized value equal to the true covariance.

The fixed-family extension uses a penalty and priorities specified given the grown labels, known $V$, and design, without further outcome dependence. It does not justify a cross-validated or manually outcome-selected penalty, a penalty based on a variance estimated from the same outcomes, a different pruning risk, subtree raising, or an outcome-selected prognostic basis. Causal interpretation requires identification in the analyzed sample and the correct reference prognosis; the Gaussian projection result alone does not provide population subgroup-ATE inference.

\subsection{A concrete change in target under GLS weighting}\label{gls:example:target}
Consider eight fixed observations with $B=[\ind,x]$ and a separate baseline split variable whose left indicator is $h$, where
\[
\begin{aligned}
x&=(-3,3,1,-4,-2,-4,3,2)\trans,\\
A&=(1,1,0,0,1,1,0,0)\trans,\\
h&=(1,1,1,1,0,0,0,0)\trans.
\end{aligned}
\]
Set the correctly specified reference prognosis to zero and let the CATEs at the design points be $(1,20,1,20,0,0,0,0)\trans$. The finite-design ordinary child-average effect difference is $21/2$ and every left CATE exceeds every right CATE. Nevertheless, the GLS projection target for this same split and unchanged mean depends on the known variance ratio:
\[
\begin{array}{c|ccc}
\rho&1/4&1&4\\\hline
\theta_s&-413/62&-35/22&119/24.
\end{array}
\]
The two treated-left contrast weights are $(87/62,-25/62)$, $(25/22,-3/22)$, and $(19/24,5/24)$, respectively. Their sum is one in every case, but the signed weights need not define a convex average. The targets follow by multiplying these weights by the treated-left CATEs $(1,20)$; treated-right effects vanish. Each contrast is obtained from~\eqref{pr:eq:contrast} with its stated known $V$. This example satisfies the mean model and can be embedded in Gaussian potential-outcome models with the stipulated variances. It demonstrates a change of estimand, not a failure of the selective inference guarantee. It also explains why the sign of a general signed projection target should not rank subgroup-average benefit.

\subsection{Two-component support geometry}
\label{gls:sec:support-shape}
The following is a GLS analogue of the two-tail geometry in Tree-Values \citep[Appendix D, Lemmas 26 and 30]{Neufeld2022}, rather than a new general selective-inference principle.

\begin{corollary}[Shape of a retained GLS path section]
\label{gls:cor:two-components}
Under the weighted construction and Assumptions~\ref{pr:ass:growth} and~\ref{pr:ass:pruning}, fix a feasible path $\gamma$ and $z\in\eta^\perp$. Its exact retained section consists of at most two intervals, allowing empty or degenerate intervals. In particular, it has at most two positive-length components and at most four finite boundary points, including any isolated points. Under Assumption~\ref{pr:ass:gaussian}, the conditional law is supported on the positive-length pieces for almost every selected $z$.
\end{corollary}
\begin{proof}
Use $S=dT$ with direction $b=v/d=V\ell$. Every selected ancestral score is constant. Each ancestral comparison therefore restricts $(aS+c)^2$ to be below a fixed nonnegative constant, strictly or weakly according to priority. Such sets, including empty sets and singletons, are intervals; their intersection is an interval $I(z)$.

For a competitor $r$ at the target, set $a_r=\ell_r\trans V\ell$ and $c_r=\ell_r\trans z$. Unit $V$-norms and duplicate removal give $|a_r|<1$. The comparison $S^2\ge(a_rS+c_r)^2$ excludes a central interval with endpoints
\[
\frac{c_r}{1-a_r},\qquad -\frac{c_r}{1+a_r}.
\]
Every nonempty exclusion contains zero. Strict comparisons include the endpoints in the exclusion. If $c_r=0$, a weak comparison excludes nothing and a strict comparison excludes only zero. Duplicate comparisons, if retained, would be redundant or make the target infeasible.

Write $\delta=\delta_\gamma(z)$. Pruning additionally excludes nothing, $\{0\}$, or $[-\sqrt\delta,\sqrt\delta]$ as $\delta$ is negative, zero, or positive. The union of all exclusions is one possibly empty central interval $J(z)$. Thus the support is $I(z)\setminus J(z)$, comprising at most two interval pieces, each possibly a singleton. This also proves the boundary bound. Scaling by $1/d$ preserves the description. Conditional absolute continuity gives the final claim.
\end{proof}

\subsection{Numerical implementation and candidate screening}
\label{supp:sec:implementation}
The exact identities and coverage statements concern the mathematical procedure in exact arithmetic. The numerical implementations first enumerate all adjacent-observation cut points and apply the stated minimum-cell rules. In the simulations, the projection coefficient on a residualized child indicator is set to zero when its computed squared norm does not exceed $10^{-12}$. Interaction candidates are retained when their computed squared residual norm is at least $10^{-16}$, corresponding to $d_s\ge10^{-8}$. Candidates whose computed unit whitened directions satisfy
\[
|u_r\trans u_s|>1-10^{-12}
\]
are represented by the first retained candidate in the fixed feature and cut-point priority order. This numerical reduction approximates the exact duplicate-score rule in Assumption~\ref{pr:ass:growth}; it is not an identity for distinct exact score matrices. The reported simulations use this numerical candidate policy. Its construction depends on the design, common basis and known variance shape, without using the observed outcomes.

The ACTG implementation instead retains the first representative of identical or complementary child partitions and does not merge merely near-parallel residual directions. Growth uses weighted QR rank checks with tolerance $10^{-9}$ and requires a residual norm of at least $10^{-8}$. These calculations use the frozen estimated arm variances described in Section~\ref{sec:actg}; freezing them does not supply the known-covariance guarantee. The Honest holdout contrasts are checked separately for rank and cell eligibility.

Every positive-length support component is retained in the probability calculations. Gaussian masses are evaluated on a log scale, and confidence endpoints are obtained by bracketing and mean inversion. The accompanying code specifies these tolerances and provides instructions for regenerating the analyses. Independent numerical checks, including high-precision checks of extreme intervals, are described in Section~\ref{sec:sim-evaluation}. These computational checks are separate from the distributional and causal assumptions of the inferential results.

\bibliographystyle{plainnat}
\bibliography{references}
\end{document}